\documentclass[lettersize,journal]{IEEEtran}
\IEEEoverridecommandlockouts
\usepackage{cite}
\usepackage{amsmath,amssymb,amsfonts}
\usepackage{algorithm}
\usepackage{algpseudocode}
\usepackage{graphicx}
\usepackage{textcomp}
\usepackage{xcolor}
\usepackage{pifont}
\usepackage{hyperref}
\usepackage{microtype}
\usepackage{booktabs} 
\usepackage{ifthen}
\usepackage{fancyhdr}
\usepackage{multirow}
\usepackage{etoolbox}
\usepackage{verbatim}
\usepackage{tikz}
\usepackage{subcaption}
\usetikzlibrary{arrows, arrows.meta, hobby}
\usetikzlibrary{external}
\def\BibTeX{{\rm B\kern-.05em{\sc i\kern-.025em b}\kern-.08em
    T\kern-.1667em\lower.7ex\hbox{E}\kern-.125emX}}

\algblock{ParFor}{EndParFor}
\algnewcommand\algorithmicparfor{\textbf{parfor}}
\algnewcommand\algorithmicpardo{\textbf{do}}
\algnewcommand\algorithmicendparfor{\textbf{end\ parfor}}
\algrenewtext{ParFor}[1]{\algorithmicparfor\ #1\ \algorithmicpardo}
\algrenewtext{EndParFor}{\algorithmicendparfor}

\newcommand{\final}{1}

\definecolor{johnColor}{rgb}{0,0.5,0.7}
\newcommand{\JDO}[1]{{\color{johnColor} JDO\@: #1}}
\ifthenelse{\equal{\final}{1}}
{
  \renewcommand{\JDO}[1]{}
}
{}

\definecolor{zhongyiColor}{rgb}{0.5,0.7,0}
\newcommand{\zhongyi}[1]{{\color{zhongyiColor} zhongyi\@: #1}}
\ifthenelse{\equal{\final}{1}}
{
  \renewcommand{\zhongyi}[1]{}
}
{}

\definecolor{louisColor}{rgb}{0.7,0.2,0}
\newcommand{\louis}[1]{{\color{louisColor} louis\@: #1}}
\ifthenelse{\equal{\final}{1}}
{
  \renewcommand{\louis}[1]{}
}
{}

\newcommand{\op}[1]{\emph{#1}}

\algnewcommand{\LineComment}[1]{\State {\color{blue}\(\triangleright\) #1}}

\renewcommand{\headrulewidth}{0pt}
\makeatletter
\patchcmd{\@makecaption}
  {\scshape}
  {}
  {}
  {}
\makeatletter
\patchcmd{\@makecaption}
  {\\}
  {.\ }
  {}
  {}
\makeatother

\newcommand{\pluseq}{\mathrel{{+}{=}}}

\newcommand{\cmark}{\ding{51}}%
\newcommand{\xmark}{\ding{55}}%

\begin{document}

\title{Towards Universal Performance Modeling for Machine Learning Training on Multi-GPU Platforms
}

\author{Zhongyi Lin, Ning Sun, Pallab Bhattacharya, Xizhou Feng~\IEEEmembership{Member, IEEE}, Louis Feng, John D. Owens~\IEEEmembership{Fellow, IEEE}
\thanks{Zhongyi Lin is with Advanced Micro Devices, Inc, San Jose, California, USA (email: \href{mailto:zhongyil@amd.com}{zhongyil@amd.com}). This work was done when he was a Ph.D. student at UC Davis and a research intern at Meta, Inc.}
\thanks{Ning Sun is with Meta, Inc, Menlo Park, California, USA (email: \href{mailto:nsun@meta.com}{nsun@meta.com}).}
\thanks{Pallab Bhattacharya is with NVIDIA, Santa Clara, California, USA (email: \href{mailto:pallabb@nvidia.com}{pallabb@nvidia.com}). This work was done when he was with Meta, Inc.}
\thanks{Xizhou Feng is with Meta, Inc, Menlo Park, California, USA (email: \href{mailto:fengx@meta.com}{fengx@meta.com}).}
\thanks{Louis Feng is with Meta, Inc, Menlo Park, California, USA (email: \href{mailto:lofe@meta.com}{lofe@meta.com}).}
\thanks{John D. Owens is with the Department of Electrical and Computer Engineering, University of California, Davis, Davis, California, USA (email: \href{mailto:jowens@ucdavis.edu}{jowens@ucdavis.edu}).}
}

\maketitle

\begin{abstract}
Characterizing and predicting the training performance of modern machine learning (ML) workloads on compute systems with compute and communication spread between CPUs, GPUs, and network devices is not only the key to optimization and planning but also a complex goal to achieve. The primary challenges include the complexity of synchronization and load balancing between CPUs and GPUs, the variance in input data distribution, and the use of different communication devices and topologies (e.g., NVLink, PCIe, network cards) that connect multiple compute devices, coupled with the desire for flexible training configurations.
Built on top of our prior work for single-GPU platforms
, we address these challenges and enable multi-GPU performance modeling\footnote{Code is open-sourced at \url{https://github.com/owensgroup/ml_perf_model}.} by incorporating (1) data-distribution-aware performance models for embedding table lookup, and (2) data movement prediction of communication collectives, into our upgraded performance modeling pipeline equipped with inter-and intra-rank synchronization for ML workloads trained on multi-GPU platforms. Beyond accurately predicting the per-iteration training time of deep learning recommendation models (DLRM) models with random configurations with a geomean error of 5.21\% on two multi-GPU platforms, our prediction pipeline generalizes well to other types of ML workloads, such as Transformer-based natural language processing (NLP) models with a geomean error of 3.00\%. Moreover, even without actually running ML workloads like DLRMs on the hardware, it is capable of generating insights such as quickly selecting the fastest embedding table sharding configuration (with a success rate of 85\%).
\end{abstract}


\section{Introduction}
\label{sec:intro}
Modern machine learning (ML) workloads tend to grow in size and computation, usually beyond the capability of one single GPU to host and train. Therefore, platforms with multiple compute devices such as GPUs are critical in undertaking training jobs for these workloads. Modern industrial recommendation models, such as DLRM~\cite{Naumov:2019:DLR}, consist of embedding tables of hundreds of gigabytes and are trained in a distributed way on hierarchical compute systems, e.g., 16 nodes $\times$ 8 GPUs, as an advanced software-hardware co-designed system~\cite{Mudigere:2021:HPD}. For large language models (LLM), the rapidly growing number of model parameters requires an equally rapid increase in the number of compute devices used to train them. Examples include GPT3~\cite{Brown:2020:LMA} (2020, 175B (parameter count, same below), 1024 NVIDIA V100 GPUs (estimated)) Megatron-Turing NLG~\cite{Smith:2022:UDA} (2022, 530B, 2240 80-GB NVIDIA A100 GPUs), PaLM~\cite{Chowdhery:2022:PSL} (2022, 540B, 6144 Google TPU v4), OPT-175B~\cite{Zhang:2022:OPT} (2022, 175B, 992 80-GB NVIDIA A100 GPUs), and LLaMA~\cite{Touvron:2023:LOA} (2023, 65B, 2048 80-GB NVIDIA A100 GPUs). For such daunting multi-GPU training jobs, ML practitioners are interested in not only the training performance (e.g., iteration time, query-per-sec (QPS), FLOPS, etc.) but also how to speed up training and use the hardware most efficiently. Characterizing the performance behavior of such multi-GPU jobs is the key to identifying and optimizing performance bottlenecks, and subsequently helping ML practitioners save development time and budget, improve ML model quality provided to users, and benefit the environment by avoiding excessive emissions of CO$_2$.

However, achieving the above goals is particularly challenging \emph{in the multi-GPU platform space} for two primary reasons:
\begin{itemize}
    \item Communication collectives, like \op{all-to-all} and \op{all-reduce} across various network media (e.g., NVLink, PCIe, network cards) and topologies that connect multiple compute devices, are essential operations in multi-GPU training and commonly the performance hotspot. The performance modeling of these operations is missing in previous single-GPU and compute-op-focused work~\cite{Justus:2018:PTC, Pei:2019:ITP, Li:2020:CAM, Liao:2020:PPA, Yu:2021:HAR, Rajagopal:2021:PAT, Lin:2022:BAP}.
    \item More importantly, the synchronization behaviors of multiple GPU streams on the same device or across multiple devices are complicated. Previous works have only inadequate or premature modeling for these behaviors, making them insufficient to account for all the GPU idle time on each rank caused by data dependencies and multi-GPU synchronization, and as a result, cannot accurately model workload execution time.
\end{itemize}

Also, even in single-GPU setups, the limitations of the performance modeling of certain operators (ops) in prior work inhibit us from extending it to a broader range of workloads. As an important example, embedding lookups in real-world DLRM pipelines have unknown input distributions and loads across batches and devices, resulting in unpredictable memory access patterns and thus difficulty in predicting their performance. Existing embedding lookup performance models are rigid and can only cover limited problem sizes. Besides, extending the performance model to other workloads such as natural language processing (NLP) models requires the support of extra element-wise or minor ops.

To summarize, the performance modeling problems of communication collectives, multi-GPU stream synchronization, embedding lookups with randomly distributed input data, and additional minor ops such as \op{layer\_norm} and \op{dropout} remained unsolved until we address them in this paper. We extend our previous performance-modeling work~\cite{Lin:2022:BAP} and inherit its critical-path-based execution trace simulation method to model the training performance of DLRM and NLP workloads on multi-GPU platforms. The novel and major contributions include:
\begin{itemize}
    \item (Section~\ref{sec:comm}) Added performance models for communication operations (\op{all-to-all} and \op{all-reduce}) by improving a basic heuristic model with straightforward and efficient sigmoid curve fitting.
    \item (Section~\ref{sec:multi_e2e}) An enhanced critical-path-based end-to-end (E2E) performance modeling algorithm.
    \item (Section~\ref{sec:el}) Improved embedding-table-lookup performance modeling with flexible lookup numbers and patterns (i.e., input data distribution) through an ML-based approach.
    \item (Section~\ref{sec:additional}) Supported extra minor ops such as \op{layer\_norm} and \op{dropout} for NLP models.
\end{itemize}
We claim that \emph{the modeling of both inter- and intra-rank synchronizations are the keys to accurately modeling ML workloads training performance on multi-GPU, or even more broadly, all types of workloads running on multi-heterogeneous-device platforms.} Therefore, the proposal of the critical-path-based end-to-end (E2E) performance modeling algorithm (second item above) is the most significant contribution of this paper. Our performance model obtains 5.21\% and 3.00\% geomean prediction errors on predicting the per-iteration performance of randomly generated industrial-scale DLRM workloads and Transformer-based NLP models based on BERT~\cite{Devlin:2019:BPO}, GPT2~\cite{Radford:2019:LMA}, and XLNet~\cite{Yang:2019:XGA}, respectively, on two different multi-GPU platforms. In addition, in a use case of choosing a (DLRM) embedding table sharding configuration without running the workloads, our performance model achieves an 85\% success rate in selecting either the fastest config or predicting the execution time with less than 10\% error, showcasing our performance model's ability to generate insights for multi-GPU training optimization.

\section{Related Works}
\label{sec:related_works}
\subsection{Performance Modeling for Both Single-Device and Distributed Training}
There is abundant prior art in predicting the execution time of per-batch training and/or inference of ML workloads on a single GPU\@. A majority of them~\cite{Justus:2018:PTC, Pei:2019:ITP, Li:2020:CAM, Liao:2020:PPA, Yu:2021:HAR, Rajagopal:2021:PAT, Lin:2022:BAP} focus on convolutional neural networks (CNNs), while some also cover NLP~\cite{Yu:2021:HAR} and recommendation models like DLRM~\cite{Lin:2022:BAP}. These works developed methodologies for accurately modeling performance on the op/kernel levels, though they did not address the challenges of multi-GPU or distributed training in general.

Early performance models for distributed training are mostly analytical and consider computation and communication separately. Yan et al.~\cite{Yan:2015:PMA} studied data and model parallelism of CNNs on CPU clusters and modeled their performance and scalability. Oyama et al.~\cite{Oyama:2016:PSO} predicted per-batch execution time for GPU-equipped supercomputers with low error based on statistics of mini-batch size and staleness for an asynchronous SGD algorithm. It also provided some generality in terms of precision and communication standards. As a use case, both approaches claimed to be able to search for the best system configurations for distributed training of CNNs. Qi et al.~\cite{Qi:2017:POT} introduced PALEO for per-layer CNN execution time estimation and demonstrated its ability to accurately model CNN training performance at scale on cloud clusters.

However, these prior successes do not extend well to modern ML workloads. Today, the implementation of ML workloads as well as the environment and setup of training clusters have evolved significantly. This outdates features such as sub-batch size (fewer than 16~\cite{Oyama:2016:PSO} vs. considerably larger sizes in modern settings, e.g., 2048 to 65536), model workload size and diversity, and hardware settings. Prior work either lacks support on GPUs~\cite{Yan:2015:PMA} or has a complex methodology that inhibits generalizing to other pipelines~\cite{Oyama:2016:PSO}. As decentralized distributed training such as DDP (distributed data-parallel) with multiple GPUs is increasingly preferred to parameter servers, modeling the communication among GPUs to account for interaction and overlap between GPU streams must also be addressed for accurate prediction. This rules out per-layer methods such as Qi et al.~\cite{Qi:2017:POT}.

Wang et al.~\cite{Wang:2019:CDL} implied a possible speed-of-light (SOL) model that the total execution time is the maximum among the time of data, computation, and communication, assuming these operations are perfectly overlapped. Inspired by this, and thanks to the detailed information we obtain from the execution trace~\cite{Sridharan:2023:CAP}, we further predict the training performance in general when these operations are not perfectly overlapped. Yang et al.~\cite{Yang:2021:PAG} proposed the computational-graph simulation-based PerfEstimator running on a single device for distributed training of CV models, with computation-communication overlap estimated by an analytical scaling factor. In our methods, this overlap is estimated by the critical-path-based method which is less error-prone especially when workloads across devices are imbalanced.

\subsection{Model Parallelism and Sharding}
As modern ML workloads have grown rapidly in size, exploiting model parallelism has become a critical tool in scaling up ML workloads. Lepikhin et al.~\cite{Lepikhin:2021:GSG} summarize the main challenges of model parallelism, which include: 1)~device underutilization due to the sequential dependency of the network and gradient-based optimization; 2)~superlinear compute cost vs.\ model size; 3)~poor infra scalability; 4)~non-trivial implementation of partitioning strategies. As an example of one increasingly common use of model parallelism, modern recommendation models such as industrial DLRMs often have hundreds of embedding lookup tables with up to tens of millions of rows, which are too large to be stored on a single GPU and thus must be sharded and dispatched to multiple GPUs. Balancing the load of these sharded tables (to achieve uniform access latency and memory usage) thus becomes a critical problem that previous researchers have tried to solve in different ways. Lui et al.~\cite{Lui:2021:UCD} first applied baseline (e.g., dimension-based, row-based, and size-based) heuristics to shard embedding tables in DLRMs. Sethi et al.\ proposed RecShard~\cite{Sethi:2022:RSF} that uses integer linear programming (ILP) to optimize the sharding problem. Zha et al.\ further improved the sharding efficiency and balance by the reinforcement-learning-based AutoShard~\cite{Zha:2022:AAE} and Dreamshard~\cite{Zha:2022:DGE}. In our paper, we demonstrate through a case study that our performance model can evaluate the performance of multiple sharding algorithms and quickly select the one that leads to the best E2E execution time for DLRM training. Notice that we are \emph{not} proposing a smart sharding algorithm here as these previous works did.


\section{Methodology}
\label{sec:methodology}
\begin{figure*}
  \begin{center}
    \centerline{\includegraphics[width=0.85\textwidth]{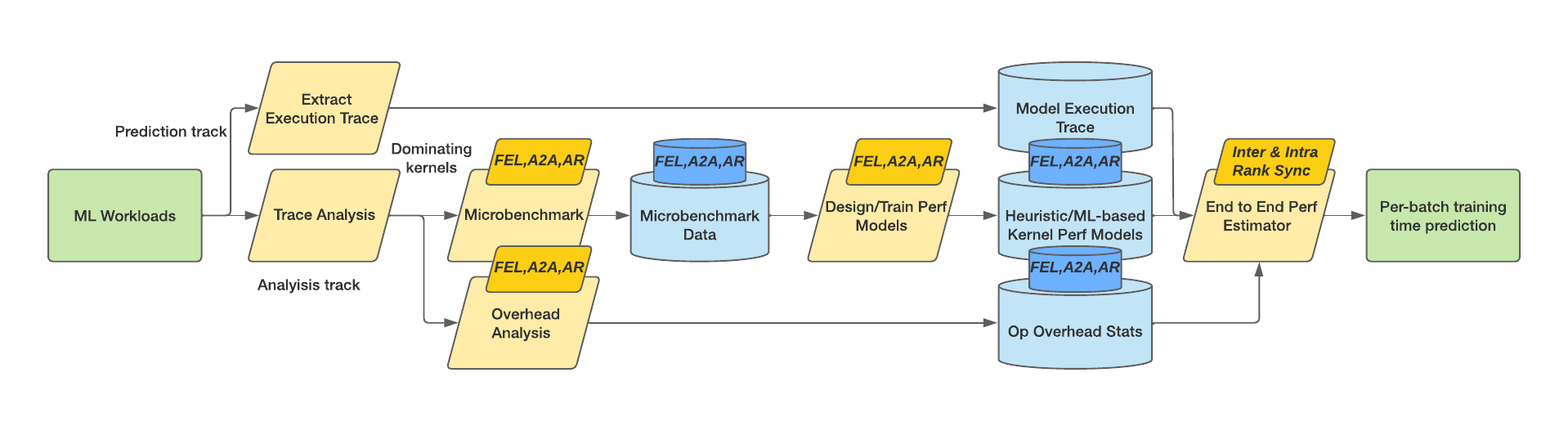}}
    \vskip -0.15in
    \caption{An overview of our prediction pipeline, based on the one proposed by Lin et al.~\cite{Lin:2022:BAP}. We mark new components with small darker shapes and italic text, such as microbenchmark and kernel-level performance models of FBGEMM embedding lookup (Section~\ref{sec:el}), all-to-all (A2A), and all-reduce (AR) (Section~\ref{sec:comm}), as well as inter- and intra-rank synchronization mechanism in the critical-path algorithm (Section~\ref{sec:multi_e2e}) to handle multi-GPU end-to-end performance prediction.}
    \label{fig:system_overview}
  \end{center}
\end{figure*}

Figure~\ref{fig:system_overview} provides an overview of our multi-GPU performance modeling pipeline with the modules that exist in our previous work~\cite{Lin:2022:BAP} marked by light blocks and extended modules of this current work marked by dark blocks and italic text. The previous work builds a solid foundation for the pipeline. In the analysis track, we profile ML workloads written with PyTorch and analyze the profiler traces to collect dominating kernels/ops. Then we collect microbenchmark data for these kernels/ops, design/train and verify performance models using this data, and extract the overhead statistics for these kernels/ops. The kernel benchmark data, kernel performance models, and overhead statistics are saved as the assets of the pipeline ((blue cylinders in Figure~\ref{fig:system_overview})). In the prediction track, the pipeline extracts the execution trace (renamed from the ``execution graph") of an input ML workload, simulates it by traversing its ops with the pipeline assets, and generates the per-iteration E2E training time prediction of the input ML workload within several seconds.

We leverage the modularity of the previous work and extend it with a focus on performance modeling on \emph{single-node multi-GPU platforms}. This makes sense for two reasons: 1) the pipeline assets are reused and shared across different input ML workloads, and 2) the E2E time prediction by simulating the execution trace of the target workload and traversing ops in it is dictated by the same critical-path-based logic that is both \emph{model-architecture-agnostic} and \emph{platform-agnostic}. In our current work, we add kernel performance models for dominating kernels/ops of ML workloads running on multi-GPU platforms and address multi-GPU execution in the E2E simulation. \emph{The updated pipeline inherits the main modules and procedures and reuses all kernel performance models (for \op{GEMM}, \op{memcpy}, \op{transpose}, etc) and the overhead estimator from the previous version, while the critical-path-based simulation algorithm is enriched to adapt to multi-GPU execution.} We focus on DLRM and Transformer-based NLP workloads in this work, but we expect our techniques can seamlessly extend and apply to other ML workloads (details discussed in Section~\ref{sec:discussion}).

\subsection{Dataset Exploration}
\label{sec:dataset_exploration}
\begin{figure}
    \centering
    \includegraphics[width=0.5\textwidth]{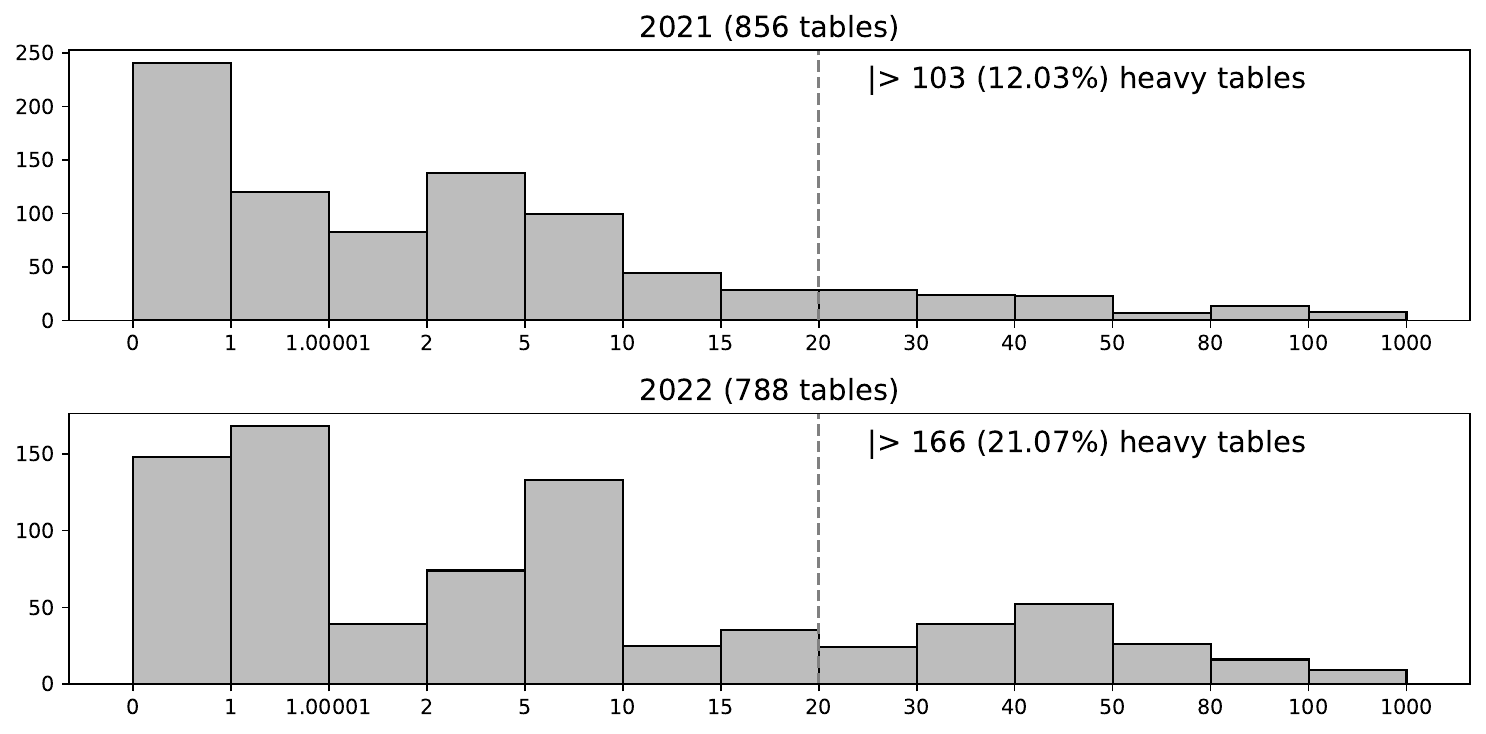}
    \caption{Histogram of average $L$ values of EL tables in the dataset. All bins but the last one are right-open.}
    \label{fig:hist}
    \vskip -0.25in
\end{figure}

The open-source DLRM dataset~\cite{Meta:2021:DOD} contains synthetic embedding lookup (EL) data that resembles the memory-access reuse pattern of Meta's production data. The 2021 data includes one batch of 65536 EL data samples from 856 tables, while the 2022 data includes one batch of 131072 samples from 788 tables. Due to the unpredictable lookup pattern in production data, the pooling factors (average number of lookups per data sample, denoted with $L$) vary significantly across all tables. This is very different from the previous work~\cite{Tulloch:2020:BEL, Lin:2022:BAP} which considers $L$ as a fixed value across data batches and even tables in the same workload. Our exploratory analysis (Figure~\ref{fig:hist}) shows that average $L$ values of tables in the dataset tend to concentrate in the range of $[0, 10)$, especially $[0, 1]$, while there are still some tables with average $L$ value spreading to a few hundred. We define tables with an average $L$ equal to greater than 20 as ``\emph{heavy tables}''. As shown in Figure~\ref{fig:hist}, there are 103 and 166 heavy tables in the 2021 and 2022 data respectively.

\begin{figure*}
  \begin{center}
    \vskip -0.3in
    \centerline{\includegraphics[width=
\textwidth]{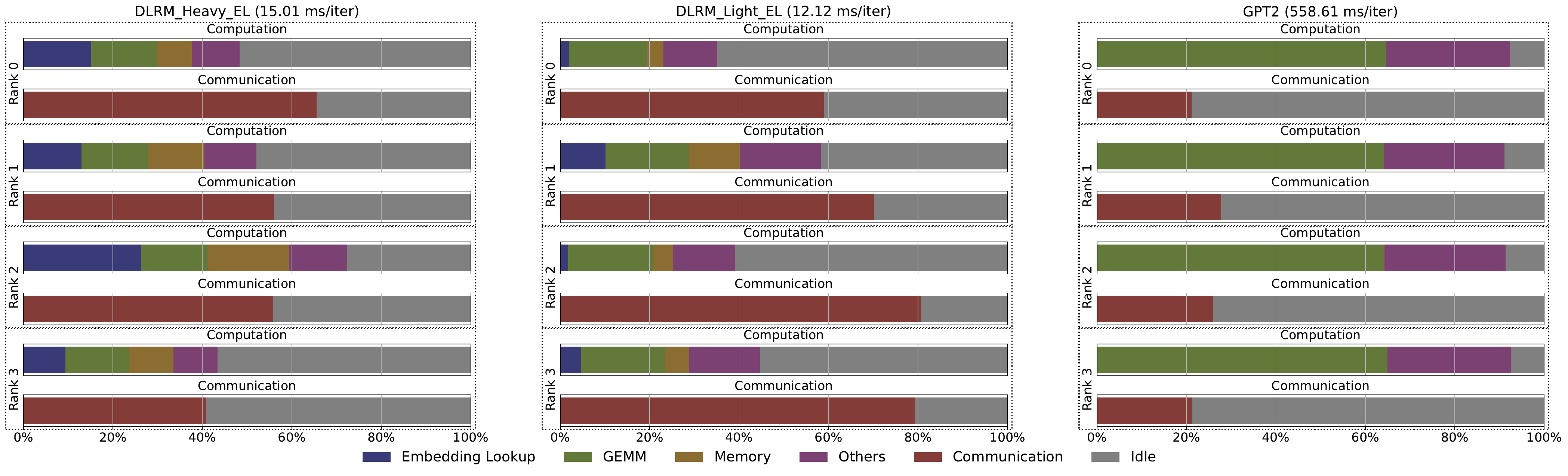}}
    \caption{Per-GPU-stream training execution time breakdown of selected ML workloads on a 4-GPU platform. The per-iteration time of each workload is provided for reference. We discuss these results in Section~\ref{sec:bench}.}
    \vskip -0.4in
    \label{fig:multi_gpu_time_breakdown}
  \end{center}
\end{figure*}

\subsection{Multi-GPU Training Benchmark and Analysis}
\label{sec:bench}
We first benchmark several ML workloads that will be further analyzed in later sections to understand the behavior of multi-GPU training performance. Figure~\ref{fig:multi_gpu_time_breakdown} shows the per-GPU-stream execution time breakdown of three workloads: \emph{DLRM\_Heavy\_EL} with a batch size of 4096 (Task\_14 in Section~\ref{sec:multi_gpu_e2e_tests}, involving heavy embedding tables only), \emph{DLRM\_Light\_EL} with a batch size of 4096 (Task\_5 in Section~\ref{sec:multi_gpu_e2e_tests}, involving random embedding tables), and \emph{GPT2} with a batch size of 64 on each rank of the 4xGV100 platform (details shown in Section~\ref{sec:evalution}). The runtime is contributed by a few groups of kernels and types: EL (forward and backward), GEMM (forward and backward), memory operations (\op{concat}, \op{copy}, \op{transpose}, etc), others (normalizations, elementwise, etc), communication (\op{all-to-all} and \op{all-reduce}), and idle time. We summarize the following insights from our benchmark:
\begin{itemize}
    \item \textbf{Communication operations could be but are not always a performance bottleneck.} Obviously, communication operations dominate the two DLRM models, mainly because in addition to \op{all-reduce} commonly caused by accumulating gradients in the backward pass, training the EL module in DLRM with model-parallelism introduces \op{all-to-all} for merging data with that from the multilayer perceptron (MLP) modules trained with data parallelism. In most pairs of streams shown here (except for \emph{DLRM\_Heavy\_EL}'s rank 2), the communication stream has a longer GPU active time than the computation stream. This means it cannot be fully overlapped by the computation, and thus similar workloads tend to thrash the communication network of the interconnected multi-GPU platform. In contrast, GPT2 is quite compute-dominated; \op{all-reduce} time is minor and can theoretically be fully overlapped.
    \item \textbf{Load balance substantially varies across the workloads.} DLRM models experience load imbalance across ranks due to the existence of EL tables with various compute and communication loads. Notice that instead of the forward and backward computation of EL itself, it is the \op{all-to-all} it introduces that dominates, \emph{especially when the EL is light}. Examples such as \emph{DLRM\_Light\_EL}'s rank 3 and 4 show that communication can occupy up to 80\% of the execution time, in which case GPU computation resources are wasted by sitting idle. In contrast, we see an almost-perfect load balance for GPT2 as it is purely DDP-trained and the load is evenly distributed.
    \item \textbf{The significance of GPU stream idle time is way more than ``being non-negligible''.} What we see in single-GPU training is that idle time is contributed mostly by CPU op calls and overheads that block the scheduling of GPU kernels; data dependencies play a less important role given there is only one stream. On the contrary, in the case of multi-GPU training data dependencies of ops lead to streams that wait on each other and thus dominate waiting time. This time cannot be predicted in the same way (i.e., statistically) as the previous work such as Lin et al.~\cite{Lin:2022:BAP} did, because the kernel execution time on each stream varies and thus the overlap of these kernels on different GPU streams is complicated and case-by-case. Therefore, to understand and \emph{simulate} how GPU streams interact and synchronize with each other is the key to accurately predicting the per-iteration time of training.
    \item \textbf{As we support more kinds of ML workloads such as (Transformer-based) NLP models, we will see an increasing number of ops that must be modeled accurately to get an accurate result.} We see that ``other'' ops (\op{layer-norm}, \op{dropout}, \op{gelu}, \op{tanh}, etc.) contribute to more than 20\% of execution time on GPT2, although each one of them only contributes a tiny amount of time every time it is executed. This might not be a surprise, given the performance of Transformer-based models has been profoundly studied these years. Techniques such as advanced layer fusion may be able to effectively optimize and reduce their execution time, but that is beyond the scope of this paper. Under our experiment settings and from the performance-modeling point of view, this observation means more ops from this category should be supported than the previous work did.
\end{itemize}

The lessons we learn guide the direction of our research, which includes the following novel components or improvements of our performance model covered in the next section:
\begin{itemize}
    \item Communication operations like \op{all-to-all} and \op{all-reduce};
    \item E2E performance modeling for multi-GPU training.
    \item Embedding table lookup with flexible lookup numbers and patterns (input data distribution);
    \item Additional ops in (Transformer-based) NLP models, such as \op{layer-norm}, \op{dropout}, \op{gelu}, and \op{tanh}, etc;
  \end{itemize}

\section{Kernel and End-to-End Performance Modeling for ML Training on Multi-GPU}
\label{sec:models}
\subsection{Communication Collective Performance Modeling}
\label{sec:comm}
\op{All-to-all} and \op{all-reduce} are two communication collectives commonly seen in multi-GPU training of recommendation workloads and contribute significantly to the per-iteration time of multi-GPU training. The difficulty of modeling the performance of these two ops is that their performance highly depends on how multiple devices are connected (e.g., network connection pattern and medium such as NVLink and PCIe), while this configuration differs from platform to platform.

\begin{figure}[t]
  \tikzstyle{line} = [draw]
  \vskip -0.05in
  \begin{subfigure}[t]{0.235\textwidth}
    \begin{tikzpicture}[scale=.65,cap=round,
        tangent/.style={%
        in angle={(180+#1)},
        Hobby finish,
        designated Hobby path=next, out angle=#1,
        }]
         \tikzset{axes/.style={}}
        \begin{scope}[style=axes]
        \draw[->] (-0.5,0) -- (5,0) node[below] {$bytes$};
        \draw[->] (0,-0.5) -- (0,4) node[right] {$\mu s$};
        \draw[-,thick] (0.1,0.3) -- (1.7,0.3);
        \draw [-,thick,use Hobby shortcut]
        (1.7,0.3) .. controls +(0:2) and +(235:1) .. (3.8,1.1);
        \draw[-,thick] (3.8,1.1) -- (4.9,2.9);
        \draw[dashed] (1.7,-0.1) -- (1.7,4);
        \draw[dashed] (3.8,-0.1) -- (3.8,4);
        \node[text width=0.1cm] at (0.9, 3.5) {1};
        \node[text width=0.1cm] at (2.6, 3.5) {2};
        \node[text width=0.1cm] at (4.3, 3.5) {3};
        \node[text width=0.1cm] at (1.7, -0.5) {$m_1$};
        \node[text width=0.1cm] at (3.4, -0.5) {$m_2$};
        \end{scope}
    \end{tikzpicture}
    \caption{Latency vs.\ message size}
    \label{fig:latency}
  \end{subfigure}
  \begin{subfigure}[t]{0.235\textwidth}
    \begin{tikzpicture}[scale=.65,cap=round,
        tangent/.style={%
        in angle={(180+#1)},
        Hobby finish,
        designated Hobby path=next, out angle=#1,
        }]
         \tikzset{axes/.style={}}
        \begin{scope}[style=axes]
        \draw[->] (-0.5,0) -- (5,0) node[below] {$bytes$};
        \draw[->] (0,-0.5) -- (0,4) node[right] {$GB/s$};
        \draw[-,thick] (0.1,0.3) -- (1.7,0.5);
        \draw [-, thick, use Hobby shortcut]
        (1.7,0.5) .. controls +(15:2) and +(180:1) .. (3.8,3.0);
        \draw[-,thick] (3.8,3.0) -- (4.7,3.0);
        \draw[dashed] (1.7,-0.1) -- (1.7,4);
        \draw[dashed] (3.8,-0.1) -- (3.8,4);
        \node[text width=0.1cm] at (0.9, 3.5) {1};
        \node[text width=0.1cm] at (2.6, 3.5) {2};
        \node[text width=0.1cm] at (4.3, 3.5) {3};
        \node[text width=0.1cm] at (1.7, -0.5) {$m_1$};
        \node[text width=0.1cm] at (3.4, -0.5) {$m_2$};
        \end{scope}
    \end{tikzpicture}
    \caption{Bandwidth vs.\ message size}
    \label{fig:bw}
  \end{subfigure}
  \vskip -0.05in
  \caption{Typical characteristic curves for data movement. X-axes are in log scale. $m_1$ and $m_2$ are boundary message sizes that separate the three regions.}
  \vskip -0.15in
  \label{fig:bw_curves}
\end{figure}
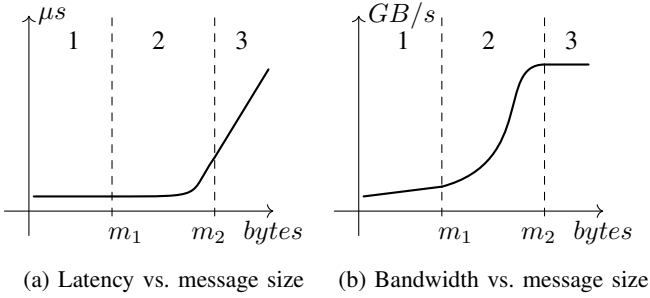

We devise a simple way to model the performance of \op{all-to-all} and \op{all-reduce} based on the message passing cost model defined by Grama et al.~\cite{Grama:2003:ITP}:
\begin{align} \label{eqn:comm_model_initial} t_{comm}=t_{cold\_start}+t_{per\_word}m_{size}\end{align} with $t_{cold\_start}$ representing the fixed overhead of data transaction, and $t_{per\_word}$ being the per-word data transaction time of the communication medium. Distilled from a comprehensive study on multi-GPU communication on platforms by Li et al.~\cite{Li:2020:EMG} and a prior performance model work (LogCA~\cite{Altaf:2017:LAH}), we make a critical observation that eventually leads to the improvement of the above model:

\begin{quote}
\emph{Regardless of the ops, network connection pattern, and network medium, as shown in Figure~\ref{fig:bw_curves} the curve of message size versus bandwidth can always be divided into three regions as the message size increases: 1) linear bandwidth with constant latency; 2) S-shape bandwidth curve with non-linear latency; 3) constant (saturated) bandwidth with linear latency.}
\end{quote}

Therefore, we model the data movement latency as a piece-wise function of message size $m$ with $m_1$ and $m_2$ being boundary sizes separating the three regions: \begin{align} \label{eqn:comm_model} t_{comm} = \left\{ \begin{array} { l l } { t_s } & { \text{ if } m \leq m_1 } \\ \text{f}(m, param) & { \text{ if } m_1 \leq m \leq m_2 } \\ t_s + \frac{m}{BW_\text{max}} & { \text{ if } m \geq m_2 } \end{array} \right. \end{align} where \begin{align} f(m, param) = \frac{\log_2^m}{10^{sigmoid(m, param)}} \end{align} and $param$ include 4 parameters $L$, $x_0$, $k$, and $b$ that define a standard sigmoid function: \begin{align} sigmoid(x) = \frac{L}{1+e^{-k*(x-x_0)}} + b. \end{align}

This method only involves 8 parameters for each op (4 for sigmoid, 2 for region boundaries, 1 for startup latency, and 1 for maximum bandwidth). It is fast and simple for both curve fitting and prediction; it is also \emph{topology-agnostic} and thus able to generalize to any communication pattern. We utilize the PARAM benchmark~\cite{Meta:2020:PB} and generate microbenchmark data to fit curves for both ops with equal message size starting from 4 bytes on each device and doubling it until exceeding the device memory. Test data is generated with random per-device message size for \op{all-to-all} and equal per-device message size for \op{all-reduce}. We calculate the input message size to the performance model as ``the maximum of (the maximum of sent/received message size per device) across all devices''.

\subsection{Multi-GPU E2E Performance Modeling}
\label{sec:multi_e2e}
As we mentioned in Section~\ref{sec:bench}, the main difference between single-GPU and multi-GPU performance modeling is that in multi-GPU scenarios, communication collectives like \op{all-to-all} and \op{all-reduce} simultaneously execute on multiple devices and trigger synchronizations and waiting across ranks and streams. This additional behavior significantly impacts runtime. This subsection describes how we simulate synchronizations and waiting during the execution analysis.

\begin{figure}
  \vskip -0.1in
  \begin{tikzpicture}[scale=0.67, every node/.style={scale=0.67}]
    \draw[dashed] (0,1.3) -- (13,1.3);
    \draw[->, -stealth] (0,-0.3) -- (12.6,-0.3);
    \node[] at (12.8,-0.3) {$t$};

    \node[] at (0.4,2.05) {$GPU_0$};
    \node[] at (1.4,1.75) {$S_{cm}$};
    \node[] at (1.4,2.35) {$S_{cp}$};

    \draw[draw=black] (1.9,2.1) rectangle ++(1,0.5) node[pos=.5] {};
    \draw[draw=black] (3,2.1) rectangle ++(0.6,0.5) node[pos=.5] {};
    \draw[draw=black] (4.2,2.1) rectangle ++(2.1,0.5) node[pos=.5] {};
    \draw[draw=black] (6.4,2.1) rectangle ++(0.7,0.5) node[pos=.5] {};
    \draw[draw=black] (7.2,2.1) rectangle ++(2.4,0.5) node[pos=.5] {};
    \draw[draw=black] (10,2.1) rectangle ++(0.2,0.5) node[pos=.5] {};
    \draw[draw=black] (10.3,2.1) rectangle ++(1.1,0.5) node[pos=.5] {};
    \draw[draw=black] (11.5,2.1) rectangle ++(1.3,0.5) node[pos=.5] {};

    \draw[draw=black] (3.7,1.5) rectangle ++(5,0.5) node[pos=.5] {};
    \draw[draw=black] (10.5,1.5) rectangle ++(2,0.5) node[pos=.5] {};
    \draw[->, -stealth, thick] (8.7,1.75) -- (10.3,2.35);

    \node[] at (0.4,0.55) {$GPU_1$};
    \node[] at (1.4,0.25) {$S_{cm}$};
    \node[] at (1.4,0.85) {$S_{cp}$};

    \draw[draw=black] (1.9,0.6) rectangle ++(1.7,0.5) node[pos=.5] {};
    \draw[draw=black] (3.7,0.6) rectangle ++(0.5,0.5) node[pos=.5] {};
    \draw[draw=black] (4.3,0.6) rectangle ++(2,0.5) node[pos=.5] {};
    \draw[draw=black] (6.4,0.6) rectangle ++(0.3,0.5) node[pos=.5] {};
    \draw[draw=black] (6.8,0.6) rectangle ++(2.2,0.5) node[pos=.5] {};
    \draw[draw=black] (9.4,0.6) rectangle ++(0.3,0.5) node[pos=.5] {};
    \draw[draw=black] (9.8,0.6) rectangle ++(1.3,0.5) node[pos=.5] {};
    \draw[draw=black] (11.2,0.6) rectangle ++(1.7,0.5) node[pos=.5] {};

    \draw[draw=black] (4.4,0) rectangle ++(4.3,0.5) node[pos=.5] {};
    \draw[draw=black] (10,0) rectangle ++(2.5,0.5) node[pos=.5] {};
    \draw[->, -stealth, thick] (8.7,0.25) -- (9.8,0.85);

    \draw[dashed,red,ultra thick] (8.7,-0.25) -- (8.7,2.75);

    \draw[densely dotted,blue,ultra thick] (9.8,-0.25) -- (9.8,1.5);
    \draw[densely dotted,blue,ultra thick] (10.3,1.2) -- (10.3,2.8);

  \end{tikzpicture}
  \vskip -0.2in
  \caption{Inter-rank synchronization (red dashed line) and intra-rank/inter-stream synchronization (blue dotted line). For simplicity, we assume two GPUs and two streams ($S_{cp}$ and $S_{cm}$, for compute and communication respectively) per GPU, while CPU op calls are omitted in the plot. Rectangles represent GPU kernels, and arrows indicate the data dependency between compute and communication kernels.}
  \label{fig:synchronizations}
\end{figure}
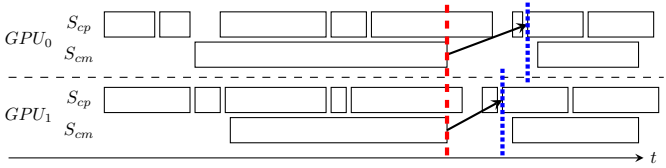

Before we dive deep into the multi-GPU E2E performance modeling algorithm, we need to first understand two types of synchronization in distributed training (depicted in Figure~\ref{fig:synchronizations}):
\begin{itemize}
    \item \emph{inter-rank synchronization} that occurs at the termination of a communication collective kernel, and
    \item \emph{intra-rank} or \emph{inter-stream synchronization} that happens at the launch of a compute/memory kernel that depends on the last communication kernel.
\end{itemize}
We expect communication kernels' launch time to differ across ranks because each rank's latency to reach these ops due to unbalanced prior loads, such as data movement and computation, is different. However, since communication kernels are synchronous, theoretically they must terminate simultaneously. In reality, they do terminate almost simultaneously with negligible variance. This is the \emph{inter-rank synchronization}, marked by the red dashed line in Figure~\ref{fig:synchronizations}. When it occurs, the time fronts of the communication streams on different ranks should be set to the same during analysis. Nevertheless, since the output data of the communication ops are used as inputs to certain successive ops, successive communication kernels, e.g., the last communication kernels on $S_{cm}$ on both ranks in Figure~\ref{fig:synchronizations}, should be launched at least after the \emph{first dependent kernel} of the previous communication kernel (marked in blue dashed lines) rather than right after the previous communication kernel. This is called \emph{intra-rank/inter-stream synchronization}. To reflect it, we set the communication stream time front to the same as the launch time of the first dependent kernel.

We claim that \emph{the modeling of both inter-rank and intra-rank synchronizations are the keys to accurately modeling ML workloads training performance on multi-GPU, or even more broadly, all types of workloads running on multi-heterogeneous-device platforms.} This is because the best modeling method without considering them is to sum the kernel time per GPU stream and take the maximum as the E2E prediction result. It might work for a few workloads dominated by one certain type of stream, e.g., GPT2 being dominated by the compute stream as shown above. However, in most cases, this will miss a large amount of GPU idle time (e.g., the gap between the red and blue lines in Figure~\ref{fig:synchronizations}) created by data dependencies and other possible waiting conditions during the execution, leading to a significant and unacceptable underestimation of execution time, i.e., up to 60\% as we show later in Section~\ref{sec:multi_gpu_e2e_tests}.

To consider these synchronizations, we need the information on data dependency between ops in the ML model/workload, which is thankfully provided by the PyTorch execution trace (ET)\@. We extend the critical-path-based algorithm proposed by Lin et al.~\cite{Lin:2022:BAP} for single-GPU performance modeling to adapt to multi-GPU scenarios by incorporating inter-rank and intra-rank synchronizations, as shown in Algorithm~\ref{alg:perf_model}. The new algorithm has a similar basic structure to the single-GPU version. Based on our observation, we assume two streams (one for compute and memory kernels and one for communication kernels) on each rank as that is the case for most ML workloads. Instead of running a single process for training time prediction on a single GPU, the new algorithm runs $N$ processes in parallel, processing $N$ EGs and predicting execution time for $N$ ranks simultaneously (line 4). In addition to CPU time and GPU active time, we also track communication stream time $T_{cm}$ and compute/memory stream time $T_{cp}$ per process (line 6). At line 7, we also initialize a variable $last\_comm\_op$ to keep track of the dependency between compute/memory ops and the last communication op. To recap, T1--T5 in line 9 represents the five types of execution overhead on the CPU of each op, which are summarized in our previous work~\cite{Lin:2022:BAP}. During the traversal of ETs, we detect intra-rank synchronization when an op that has data dependency on the last communication op or a new communication op is encountered (lines 12--15); when the current op is a communication op, we do inter-rank synchronization after processing all kernels of it (lines 29--31) and update $last\_comm\_op$ for the next iteration (lines 36--38). At the end (lines 40--41), we calculate the local total time and GPU active time and do an \op{all\_gather} operation across all ranks to collect their values, the maximums of which are returned as the predicted total time and GPU active time.

\setlength{\textfloatsep}{0pt}
\begin{algorithm}[t]
  \caption{Multi-GPU E2E Training Performance Model.}
  \label{alg:perf_model}
  \begin{algorithmic}[1]
    \State {\bfseries Input:} $ET_{0,1,\ldots N-1}$ of an ML workload trained on a single-node N-GPU platform, one execution trace per rank; Kernel performance models $\{M\}$; Overhead statistics $Ov$.
    \State {\bfseries Output:} Predicted per-batch training time $T$.

    \State Spawn N processes $P_{0,1,\ldots,N-1}$.
    \ParFor{$i\gets 0, N-1$}
        \State Initialize $cpu\_time = 0$ and $gpu\_time = 0$ for $P_i$.
        \State Initialize communication stream time $T_{cm}$ and compute stream time $T_{cp}$ for process $P_i$.
        \State Initialize $last\_comm\_op$ as none.
        \For{each $op$ in $EG_{i}$}
            \State Look up $T1, T2, T3, T4, T5$ from $Ov$ for $op$.
            \State Identify current stream $s$ (as $cm$ or $cp$).
            \LineComment{Intra-rank synchronization}
            \If{$op$ depends on $last\_comm\_op$, or $op$ is an communication op}
                \State Set $last\_comm\_op$ to none.
                \State Synchronize $T_{cm}$ and $T_{cp}$ for process $P_i$.
            \EndIf
            \State $cpu\_time \pluseq T1$
            \If{$op$ has kernel calls}
                \State $cpu\_time \pluseq T2$
                \For{each kernel call $k$ under $op$}
                    \State Predict kernel time $T_k$ with $M$
                    \State $T_s = \max(T_s + 1,cpu\_time + T4 / 2) + T_k$
                    \State Update $gpu\_time$ with $T_s$ and $T_k$.
                    \State $cpu\_time \pluseq T4$
                    \If{$k$ is not the last kernel}
                        \State $cpu\_time \pluseq T5$
                    \EndIf
                \EndFor
                \LineComment{Inter-rank synchronization}
                \If{$op$ is a communication op}
                    \State Synchronize $T_{cm}$ across all processes.
                \EndIf
                \State $cpu\_time \pluseq T3$
            \Else
                \State $cpu\_time \pluseq T5$
            \EndIf
            \If{$op$ is a communication op}
                \State $last\_comm\_op = op$
            \EndIf
        \EndFor
        \State $T = \max(T_{cm}, T_{cp}, cpu\_time)$
        \State Synchronize $T$ and $gpu\_time$ across all processes and return their maximums.
    \EndParFor
    \end{algorithmic}
\end{algorithm}

\subsection{Embedding Lookup Kernel Modeling}
\label{sec:el}
The performance modeling of \op{EL}, an indispensable and commonly dominating~\cite{Lin:2022:BAP} operator in many recommendation models including DLRM, is challenging in a few aspects. First, the dimensions and parameters of every single embedding table, specifically the \emph{number of embeddings} ($E$), the \emph{embedding dimension} ($D$), and the \emph{pooling factor} ($L$) can all be different in real workloads and data. Second, the input data distribution of each table varies considerably across each batch. Both factors significantly increase the difficulty of applying a heuristic-based performance model to the \op{EL} op, since it is almost impossible to estimate the L2 cache hit rate and data movement accurately and thus predict the latency of the op. Previous work such as Lin et al.~\cite{Lin:2022:BAP} only considers modeling the performance of the case when $D$, $E$, and $L$ are all fixed for each table with uniform input data distribution, making their solution insufficient to deal with real-world workloads and data.

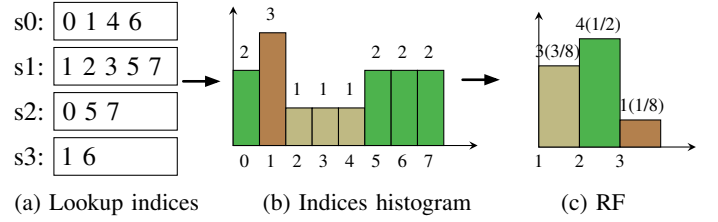
\begin{figure}
  \tikzstyle{line} = [draw]
  \begin{subfigure}[t]{0.15\textwidth}
    \tikzstyle{block} = [draw, rectangle, text width = 4em, minimum height = 5mm, node distance = 5em]
    \begin{tikzpicture}[scale=.5,cap=round]
        \node at (-1,3.6) {s0:};
        \node at (-1,2.4) {s1:};
        \node at (-1,1.2) {s2:};
        \node at (-1,0) {s3:};

        \node [block] at (1.3,3.6) {0 1 4 6};
        \node [block] at (1.3,2.4) {1 2 3 5 7};
        \node [block] at (1.3,1.2) {0 5 7};
        \node [block] at (1.3,0) {1 6};
        
        \draw[-{Triangle[width=4pt,length=6pt]}, line width=1pt] (3.1,2) -- (4.1,2);
    \end{tikzpicture}
    \caption{Lookup indices}
    \label{fig:indices}
\end{subfigure}
\hfill
\begin{subfigure}[t]{0.2\textwidth}
    \tikzstyle{block} = [draw, rectangle, text width = 4em, minimum height = 5mm, node distance = 5em]
    \begin{tikzpicture}[scale=.5,cap=round,axisline/.style={-stealth},font=\scriptsize]
        \draw [axisline] (0,0) -- (6.5,0) node[right]{};
        \draw [axisline] (0,0) -- (0,3.7) node[right]{};

        \draw[draw=black,fill=green!40!gray] (0,0) rectangle ++(0.7,2);
        \node at (0.3,2.5) {2};
        \draw[draw=black,fill=orange!40!gray] (0.7,0) rectangle ++(0.7,3);
        \node at (1,3.5) {3};
        \draw[draw=black,fill=yellow!40!gray] (1.4,0) rectangle ++(0.7,1);
        \node at (1.7,1.5) {1};
        \draw[draw=black,fill=yellow!40!gray] (2.1,0) rectangle ++(0.7,1);
        \node at (2.4,1.5) {1};
        \draw[draw=black,fill=yellow!40!gray] (2.8,0) rectangle ++(0.7,1);
        \node at (3.1,1.5) {1};
        \draw[draw=black,fill=green!40!gray] (3.5,0) rectangle ++(0.7,2);
        \node at (3.8,2.5) {2};
        \draw[draw=black,fill=green!40!gray] (4.2,0) rectangle ++(0.7,2);
        \node at (4.5,2.5) {2};
        \draw[draw=black,fill=green!40!gray] (4.9,0) rectangle ++(0.7,2);
        \node at (5.2,2.5) {2};

        \node at (0.3,-0.4) {0};
        \node at (1,-0.4) {1};
        \node at (1.7,-0.4) {2};
        \node at (2.4,-0.4) {3};
        \node at (3.1,-0.4) {4};
        \node at (3.8,-0.4) {5};
        \node at (4.5,-0.4) {6};
        \node at (5.2,-0.4) {7};

        \draw[-{Triangle[width=4pt,length=6pt]}, line width=1pt] (6.1,1.75) -- (7.1,1.75);
    \end{tikzpicture}
    \caption{Indices histogram}
    \label{fig:idx_hist}
\end{subfigure}
\hfill
\begin{subfigure}[t]{0.1\textwidth}
    \tikzstyle{block} = [draw, rectangle, text width = 4em, minimum height = 5mm, node distance = 5em]
    \begin{tikzpicture}[scale=.36,cap=round,axisline/.style={-stealth},font=\scriptsize]
        \draw [axisline] (0,0) -- (5.8,0) node[right]{};
        \draw [axisline] (0,0) -- (0,5) node[right]{};

        \draw[draw=black,fill=yellow!40!gray] (0,0) rectangle ++(1.5,3);
        \node at (0.7,3.5) {3(3/8)};
        \draw[draw=black,fill=green!40!gray] (1.5,0) rectangle ++(1.5,4);
        \node at (2.2,4.5) {4(1/2)};
        \draw[draw=black,fill=orange!40!gray] (3,0) rectangle ++(1.5,1);
        \node at (3.8,1.5) {1(1/8)};

        \node at (0,-0.5) {1};
        \node at (1.5,-0.5) {2};
        \node at (3,-0.5) {3};
    \end{tikzpicture}
    \caption{RF}
    \label{fig:rf_sub}
\end{subfigure}
  \caption{Generation of reuse factors (RF) from EL indices. (a) is an example of lookup indices of a batch of 4 samples on one embedding table. In (b), the x-axis is the indices and the y-axis is the count of these indices. In (c), the x-axis is the count of accesses and the y-axis is the number of such indices.}
  \label{fig:rf}
\end{figure}

To address this issue, we adopt a straightforward and clear expression of describing \op{EL}'s input data distribution using \emph{reuse factors} (RF), as introduced by Meta's open-source DLRM dataset~\cite{Meta:2021:DOD}. Figure~\ref{fig:rf} shows an example of generating reuse factors (RF) from EL indices. First, the histogram of per-table index counts in a batch of input data is calculated ((a) to (b)). Then, the ``histogram of histogram'', i.e., how many indices are accessed once, twice, thrice, etc., is calculated, which is thus the RF of this batch on a certain table ((b) to (c)). In practice, the bins for this step have sizes of exponents of 2, e.g., $[0, 1)$, $[1, 2)$, $[2, 4)$, $[4, 8)$, etc. Finally, the counts of every bin are normalized into the range of $[0, 1]$. In this way, the distribution of lookup indices is described as by what probability any row of a table is hit by a certain range of times, such as $2^m$ to $2^n$ times where $m$ and $n$ are both integers and $m < n$. Meta's DLRM dataset provides the RF of 17 bins for each data file, and we follow this convention in our experiments. The advantages of RF are its low overhead in terms of processing time and parameter size. The calculation is relatively simple, and thus it is trivial to obtain RF values of each batch of input data on the fly before the training of a batch is started. Because each table has 17 RF values and the number of tables residing on each rank is usually limited, e.g., around 10, the total number of RF values for each batch of input data remains acceptable.

\begin{figure}
    \centering
    \includegraphics[width=0.5\textwidth]{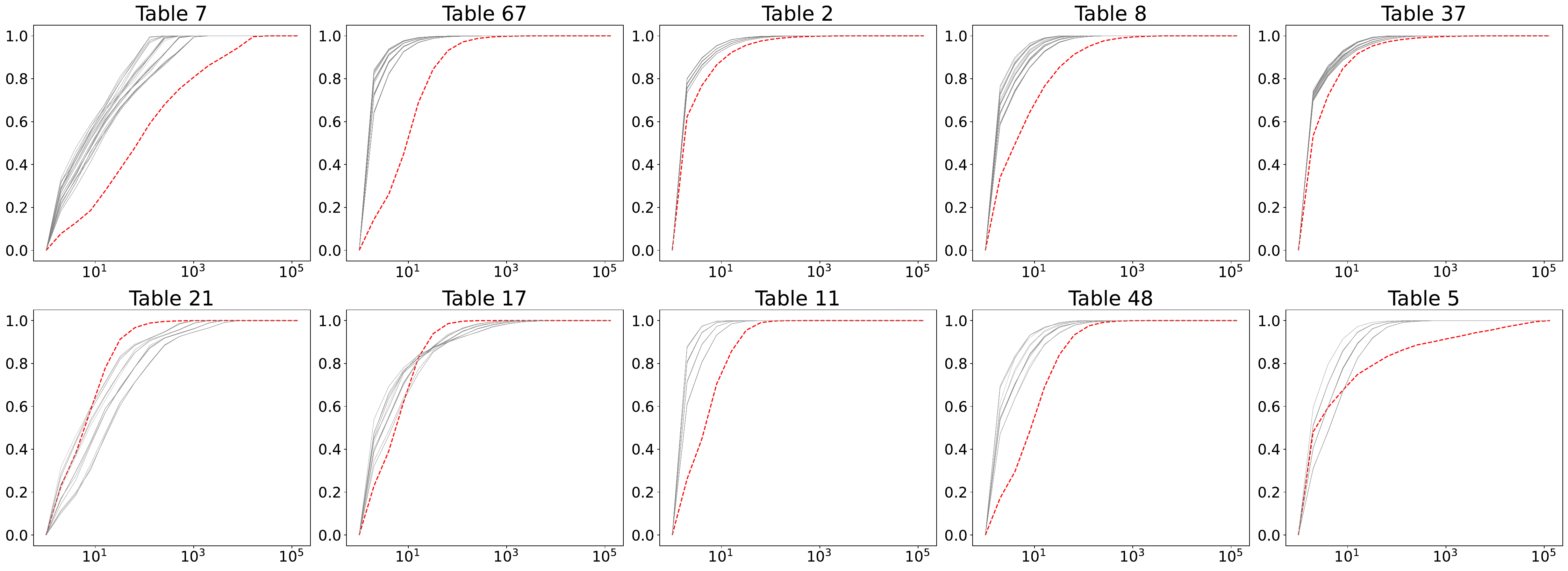}
    \caption{RF values of the top 10 most visited tables among 6000 random samples from the DLRM open-source dataset for the EL microbenchmark. Red dashed lines represent the cumulative distribution function (CDF) of the overall RF values of each table in the whole dataset, while each thin gray line represents the CDF of RF values of a sample that visits a certain table. It can be observed that the data distribution (represented by RF values) of each data batch is different and sometimes distant from that of the whole dataset. Compared to overall RF values, the RF values of a data batch tend to concentrate towards 0 along the X axis.}
    \label{fig:rfs}
\end{figure}

We adapt the cost model proposed by Zha et al.~\cite{Zha:2022:DGE} and use it in real-time performance modeling of \op{EL}\@. Zha et al.\ create a multi-head MLP cost model to simultaneously predict the cost of the forward time, backward time, and communication time of a sharding scheme of \op{EL} as a whole on multiple devices. Our method is different in two aspects. 1) We handle these three constituent times separately in the granularity of ops to accommodate our robust E2E algorithm shown in Section~\ref{sec:multi_e2e}. This better simulates and models the actual execution since considering these three times as a whole misses the opportunity to explain the computation-communication overlaps (described in Section~\ref{sec:multi_e2e}). 2) Instead of using RF values of each table, we use RF values of each \emph{batch} for better prediction based on the reason stated in the last paragraph. We construct our MLP performance model resembling one of the heads in Zha et al.'s cost model and match those used for other kernels for convenient training and inference. To enable \op{EL}'s execution incorporating flexible $D$, $E$, and $L$, we use a popular, efficient, and flexible batched-EL implementation provided by FBGEMM~\cite{Khudia:FEH:2021}, an open-source high-performance kernel library for training and inference on both CPUs and GPUs. We first create a microbenchmark dataset by randomly sampling tables and batches (e.g., 10 tables out of 856, and a batch of 1024 lookups out of 65536) from the DLRM open-source dataset, and make sure that the size of the sampled tables in each sample will not exceed the DRAM memory size of the selected GPU\@. Although Zha et al. trained the cost model with RF values of \emph{the whole dataset}, it is improper to do so for real-time performance modeling because RF values tend to spread in a larger range. This fact is easy to understand intuitively. Suppose the batch size is small, the number of times each row in a table is hit tends to concentrate close to 0 because there are few indices in the batch. If we consider the whole dataset as a huge batch (e.g., size 65536 for the DLRM dataset we use), there are likely some rows in a table to be hit many times, such as 500 times, which could never appear with a small batch size such as 256. In a word, these two distributions are different (as shown in Figure~\ref{fig:rfs}), and thus the distribution of the whole dataset is not representative of a batch of data sampled from it. Therefore, we calculate the RF values of \emph{each batch} and store them together with the execution time as part of the training data. Finally, we randomly split the microbenchmark result dataset into a training set and a test set with a split factor of 0.8, and use them to train and validate the performance models.

\subsection{Additional Ops Performance Modeling Support}
\label{sec:additional}
To support Transformer-based NLP models, We also add kernel performance models for additional ops including \op{layer\_norm}, \op{dropout}, and element-wise ops such as \op{gelu} and \op{tanh}. By applying principles mentioned in our previous work~\cite{Lin:2022:BAP}, ML-based models suit \op{layer\_norm} and \op{dropout} for their hidden kernel implementation detail in PyTorch, while the roofline model suits element-wise ops for their simplicity. Therefore we predict the latency of \op{layer\_norm}, \op{dropout} with ML-based models trained with microbenchmark data of PyTorch ops for and predict \op{gelu} and \op{tanh} using the roofline model.

\section{Evaluation and Analysis}
\label{sec:evalution}
\begin{figure}
    \centering
    \begin{subfigure}[t]{0.23\textwidth}
    \centering
    \begin{tikzpicture}[
        line width=0.6pt,
        every node/.style={circle, draw, fill, minimum size=6pt, inner sep=0pt, font=\scriptsize\bfseries}]
        \pgfsetxvec{\pgfpoint{0.9cm}{0.0cm}}
        \pgfsetyvec{\pgfpoint{0.0cm}{0.9cm}}
        \foreach \point / \id / \angle in {
            (0,1)/GPU1/270,
            (0,3)/GPU0/270,
            (2,3)/GPU3/270,
            (2,1)/GPU2/270
        } {
            \node (\id) at \point [label=\angle:\id] {};
        }
        \draw (GPU0) -- (GPU1);
        \draw (GPU2) -- (GPU3);
        \draw (GPU0) -- (GPU2) [dashed];
        \draw (GPU0) -- (GPU3) [dashed];
        \draw (GPU1) -- (GPU2) [dashed];
        \draw (GPU1) -- (GPU3) [dashed];
    \end{tikzpicture}
    \caption{4xGV100}
    \label{fig:4gv100}
\end{subfigure}
\hfill
\begin{subfigure}[t]{0.23\textwidth}
    \centering
    \begin{tikzpicture}[
        line width=0.6pt,
        every node/.style={circle, draw, fill, minimum size=6pt, inner sep=0pt, font=\scriptsize\bfseries}]
        \pgfsetxvec{\pgfpoint{0.9cm}{0.0cm}}
        \pgfsetyvec{\pgfpoint{0.0cm}{0.9cm}}
        \foreach \point / \id / \angle in {
            (0,1)/GPU1/270,
            (0,3)/GPU0/270,
            (2,3)/GPU3/270,
            (2,1)/GPU2/270
        } {
            \node (\id) at \point [label=\angle:\id] {};
        }
        \draw[line width=1.6pt] (GPU0) -- (GPU1);
        \draw[line width=1.6pt] (GPU2) -- (GPU3);
        \draw[line width=1.6pt] (GPU0) -- (GPU2);
        \draw[line width=1.6pt] (GPU0) -- (GPU3);
        \draw[line width=1.6pt] (GPU1) -- (GPU2);
        \draw[line width=1.6pt] (GPU1) -- (GPU3);
    \end{tikzpicture}
    \caption{4xA100}
    \label{fig:4a100}
\end{subfigure}
    \caption{Communication topologies of the two multi-GPU platforms used in our experiments. Thin lines: 4 NVLinks (NV4); thick lines: 12 NVLinks (NV12); dashed lines: PCIe.}
    \vskip 0.1in
    \label{fig:topology}
\end{figure}
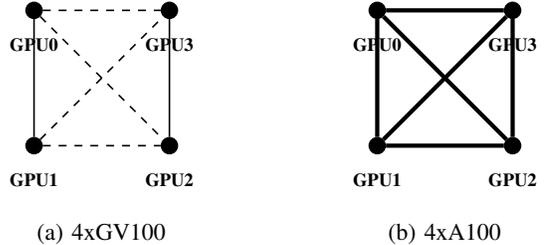

We evaluate our kernel and E2E performance models with PyTorch v2.0, FBGEMM v0.4.1, CUDA 11.7, and Python 3.9. The performance model of FBGEMM's embedding lookup kernel is evaluated on single NVIDIA GV100 and A100 (40~GB) GPUs, while that of both communication kernels and multi-GPU E2E are assessed on two multi-GPU platforms, including 4xGV100 equipped with 48-core Intel(R) Xeon(R) Gold 6146 CPU @ 3.20~GHz, and 4xA100 equipped with GCP's a2-highgpu-4g with 48-vCPU\@. Figure~\ref{fig:topology} shows the GPU communication topologies of these two platforms. We use data sampled from the DLRM open-source dataset~\cite{Meta:2021:DOD} as both microbenchmark data for FBGEMM embedding lookup kernel performance model training and verification, and input data for embedding lookup of the DLRM models in E2E tests. Multiple pieces (.pt files) of the dataset are merged for later use. The multi-GPU E2E evaluation in this work covers DLRM model training (code adapted from \url{https://github.com/facebookresearch/dlrm}) and fine-tuning of natural language processing (NLP) models such as BERT~\cite{Devlin:2019:BPO}, GPT2~\cite{Radford:2019:LMA}, and XLNet~\cite{Yang:2019:XGA} (all model implementations called from HuggingFace's Transformers library~\cite{HuggingFace:2020:TSO}). We use PyTorch's DistributedDataParallel (DDP) in distributed training of ML workloads in our experiments for no GIL contention and less model replication and data movement overheads~\cite{Meta:2023:CBD}. In addition, we manually insert a barrier at the beginning of each batch for better profiling of communication ops; we set the bucket size of gradient bucketing~\cite{Li:PDE:2020} to the default value (25) for all our experiments, as we observe that varying it only causes a negligible performance change.

\begin{table}
    \caption{Prediction error of FBGEMM embedding lookup, all-to-all, and all-reduce kernel performance models. Abbreviations: \textbf{A2A} (all-to-all), \textbf{AR} (all-reduce), \textbf{ELF} (embedding lookup forward), \textbf{ELB} (embedding lookup backward).}
    \label{kernel_prediction_error}
    \centering
    \begin{tabular*}{0.475\textwidth}{l|ccc|ccc}
        \toprule
        \multirow{2}{*}{Kernel} & \multicolumn{3}{c}{4xGV100} & \multicolumn{3}{c}{4xA100} \\
        & GMAE & MAPE & std & GMAE & MAPE & std\\
        \midrule
        A2A & 6.28\% & 9.42\% & 7.76\% & 5.25\% & 7.14\% & 4.72\% \\
        AR & 6.35\% & 9.17\% & 8.23\% & 4.98\% & 6.77\% & 4.11\% \\
        ELF & 4.37\% & 7.11\% & 7.44\% & 5.64\% & 9.17\% & 9.52\% \\
        ELB & 3.08\% & 4.42\% & 3.39\% & 3.63\% & 5.44\% & 4.33\% \\
        \bottomrule
    \end{tabular*}
\end{table}

\begin{figure}
	\centering
     \begin{subfigure}[t]{0.24\textwidth}
         \centering
         \includegraphics[width=\textwidth]{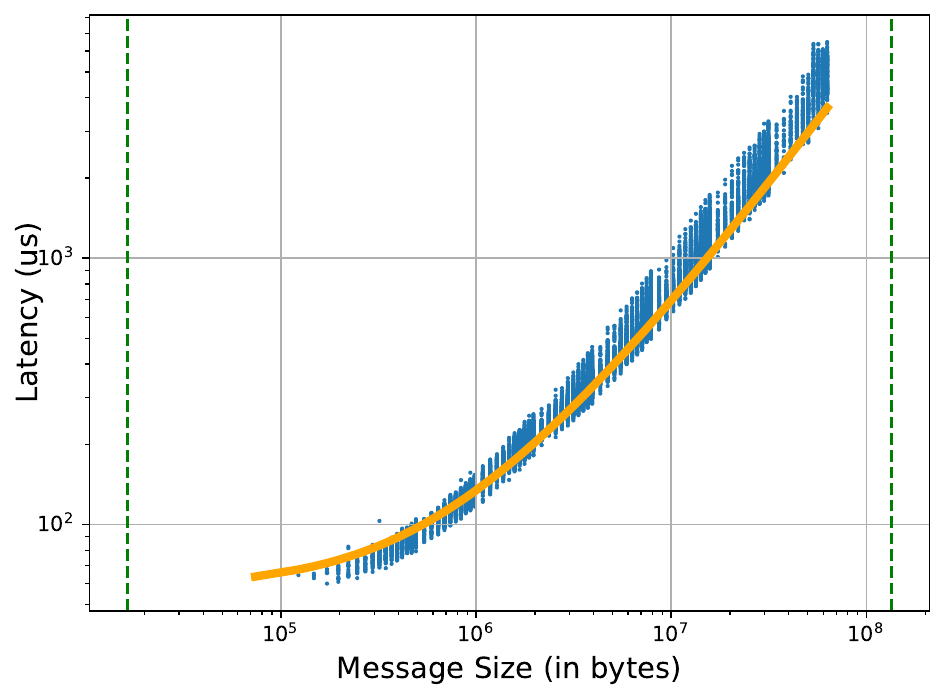}
         \caption{4xGV100}
         \label{fig:curve_4xgv100}
     \end{subfigure}
     \hfill
     \begin{subfigure}[t]{0.24\textwidth}
         \centering
         \includegraphics[width=\textwidth]{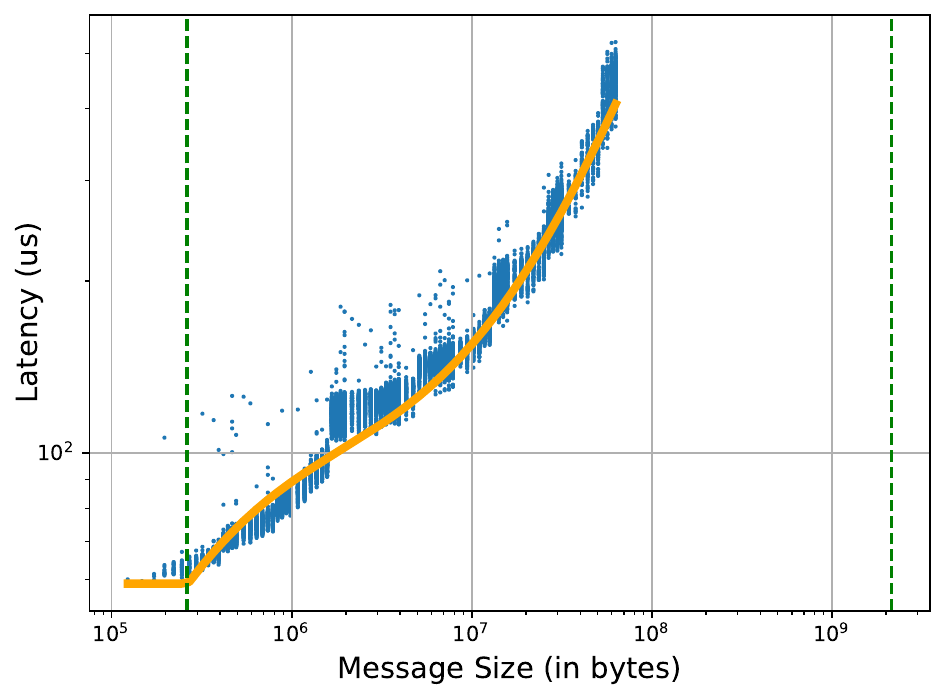}
         \caption{4xA100}
         \label{fig:curve_4xa100}
     \end{subfigure}
    \caption{Fitted curves for \op{all-to-all} benchmark data on both platforms. Message size section boundaries are plotted as green vertical dashed lines.}
    \label{fig:collective_bw}
\end{figure}

\begin{table}
  \vskip 0.1in
  \caption{Statistics of DLRM E2E time prediction errors across two multi-GPU platforms.}
  \label{table:error_stats}
  \centering
  \setlength{\tabcolsep}{1.7pt}
    \begin{small}
      \begin{tabular*}{0.485\textwidth}{ccc|ccc|ccc}
        \toprule
        \multicolumn{3}{c}{\bf{Overall}} & \multicolumn{3}{c}{4xGV100} & \multicolumn{3}{c}{4xA100} \\
        g.m. & min & max & g.m. & min & max & g.m. & min & max \\
        \midrule
        \bf5.21\% & 0.05\% & 19.38\% & 5.60\% & 0.27\% & 19.38\% & 4.85\% & 0.05\% & 17.87\%\\
        \bottomrule
      \end{tabular*}
    \end{small}
\end{table}

\subsection{Kernel Performance Modeling}
We provide prediction errors in two metrics: GMAE (geometric mean absolute error) to reduce the impact of outliers that is common when measuring E2E latency, and MAPE (mean absolute percentage error) as an intuitive measurement of deviation as an (always positive) percentage. We obtain less than 10\% both GMAE and MAPE for all kernel performance models shown in Table~\ref{kernel_prediction_error}. Particularly, the adjustment of \op{all-to-all} message size yields a low prediction error of latency, implying that the operation is bounded by the biggest per-device data bulk sent from or received by one certain device. Specifically, we present the fitted curves for \op{all-to-all} benchmark data on both platforms in Figure~\ref{fig:collective_bw}. We observe that the practical problem size for \op{all-to-all} in DLRM workloads lies in section 2 (the transitional section), which justifies our improvement on Equation~\ref{eqn:comm_model_initial}.

\begin{figure*}[ht!]
  \centering
    \begin{subfigure}[t]{0.95\textwidth}
        \centering
        \includegraphics[width=\textwidth]{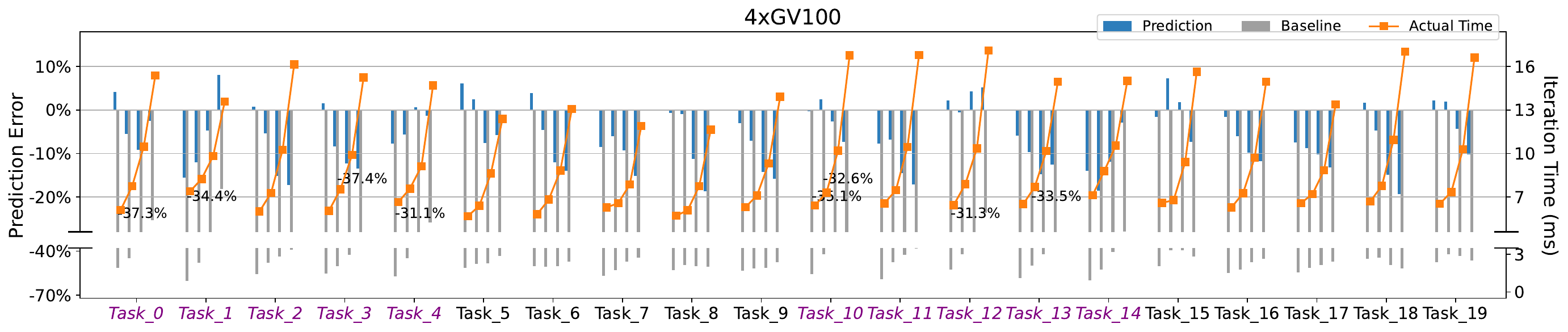}
        \label{fig:multi_e2e_gv100}
    \end{subfigure}
    \vskip -0.2in
    \begin{subfigure}[t]{0.95\textwidth}
        \centering
        \includegraphics[width=\textwidth]{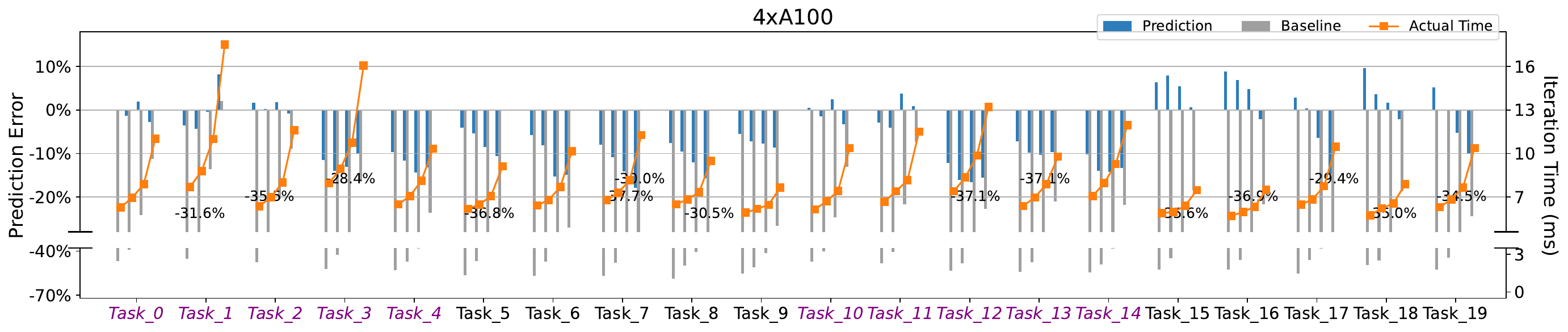}
        \label{fig:multi_e2e_a100}
    \end{subfigure}
    \vskip -0.2in
    \caption{Prediction, baseline, and reference of multi-GPU training performance of DLRMs on two multi-GPU platforms. Notice that the ``Task\_x''s on different platforms do not have the same embedding table configuration as the random task generation is platform-specific. Tasks 0--9 and 10--19 on each platform are from the 2021 and 2022 parts of the dataset, respectively. Heavy tasks are marked with purple color and italics. Percentage figures are the error of bars ending in the broken area.}
    \label{fig:dlrm_prediction}
  \vskip -0.2in
\end{figure*}

\subsection{Multi-GPU E2E Performance Modeling}
\subsubsection{DLRM on Multi-GPU Platforms}
\label{sec:multi_gpu_e2e_tests}
In our experiments to predict the multi-GPU training performance of DLRM models, we first generate 20 tasks with embedding tables randomly sampled from the DLRM open-source dataset. Among these 20 tasks, each of the 2021 and 2022 parts of the dataset contributes 5 heavy tasks (i.e., all sampled tables are heavy) and 5 normal tasks (i.e., tables can either be heavy or light). The total number of embedding tables per task is within the range of $(0.7\sim 1.3) \times \#\text{GPU} \times 13$ to resemble real-world workloads. To prevent out-of-memory errors, we limit the total memory footprint of all embedding tables on each rank to less than 80\% DRAM size of the GPU for each task. We set the embedding table sharder to size\_lookup\_greedy based, i.e., the cost of each table is estimated as $L \times D \times \log(E)$. For each training iteration in the E2E test of each task, we randomly sample a mini-batch of data (with batch size set to 512/1024/2048/4096) corresponding to each table selected in the task from the dataset and distribute it to each rank based on the sharding scheme. The overhead statistics (mean latency, etc) of PyTorch ops are aggregated from all collected traces and shared by \emph{all} tested workloads. The actual/predicted time is measured/calculated by averaging over 30 iterations. It takes about a day to run the benchmark and analysis for all 80 (20 workloads $\times$ 4 batch sizes) tasks, while it takes less than 20 minutes to predict their performance.

We present the statistics of all E2E performance modeling tests on the two multi-GPU platforms in Table~\ref{table:error_stats}, and the prediction error and reference time of each task in Figure~\ref{fig:dlrm_prediction}. The baseline result compared with our prediction in Figure~\ref{fig:dlrm_prediction} is given by the maximum sum of kernel active time of each GPU stream. We can see that this baseline prediction, yielding error values higher than 60\%, is insufficient to be used as the predicted E2E time per iteration because no idle or waiting time caused by data dependency between the streams is considered. Instead, our enhanced algorithm can accurately predict both normal and heavy tasks with high accuracy, with an overall geomean prediction error of 5.21\%. Most prediction results on both devices underestimate the actual time, possibly because the communication time dominates the per-iteration time when batch size is big, so the syncing and waiting time among all ranks also increases and contributes to the per-iteration time. The remaining tests that overestimate might be explained by the overestimation of GPU idle time caused by CPU overheads when the workload is latency-bound. Device-wise, the behavior of prediction errors, such as geomean/minimum/maximum error and trends when the batch size changes, does not deviate much, which justifies the consistency and stability of our prediction algorithm across platforms.

\begin{figure}
    \centering
    \begin{subfigure}[h]{0.475\textwidth}
        \centering
        \includegraphics[width=\textwidth]{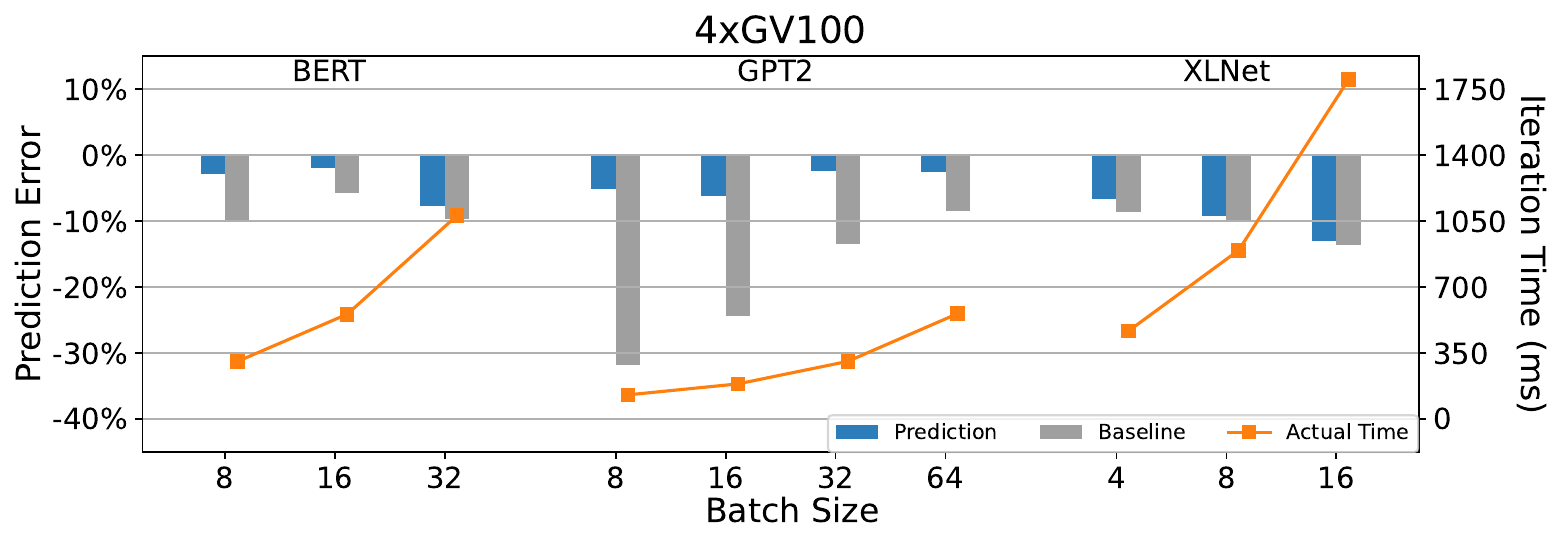}
        \label{fig:multi_e2e_nlp_gv100}
    \end{subfigure}
    \vskip -0.15in
    \begin{subfigure}[h]{0.475\textwidth}
        \centering
        \includegraphics[width=\textwidth]{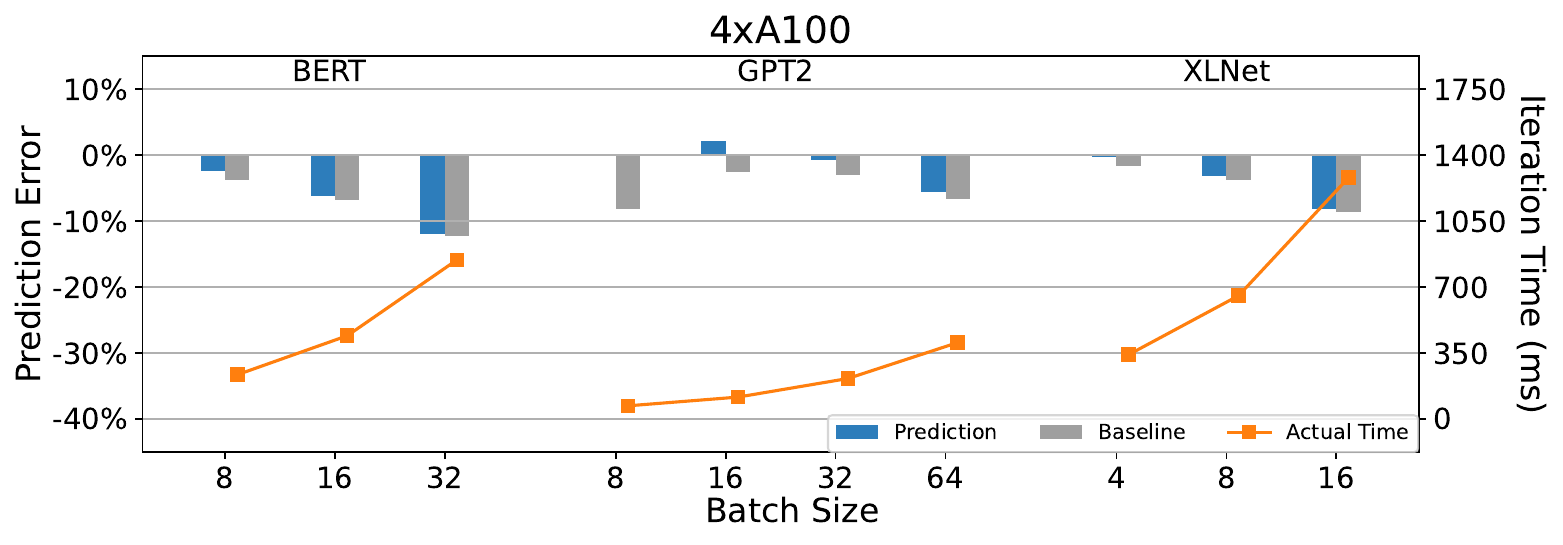}
        \label{fig:multi_e2e_nlp_a100}
    \end{subfigure}
    \vskip -0.2in
    \caption{Prediction, baseline, and reference of multi-GPU training performance of BERT, GPT2, and XLNet on two multi-GPU platforms. Batch sizes greater than 32 on BERT and 16 on XLNet result in out-of-memory errors and these tests are thus skipped.}
    \label{fig:nlp_prediction}
\end{figure}

\subsubsection{NLP Models Performance Modeling}
We also test our E2E performance model on Transformer-based NLP models including BERT, GPT2, and XLNet. In Figure~\ref{fig:nlp_prediction}, we see that the absolute prediction errors are less than 10\% in all tests except two, which are slightly larger. The geomean prediction error of all the presented tests is 3.00\%. Also, the variance of the prediction errors is lower than that of DLRM workloads. The reasons are 1) these NLP models are compute (GEMM) dominated with the communication stream being well-overlapped by the compute stream, and 2) the loads across devices are more balanced than the DLRM workloads, thus intra-rank and inter-rank synchronizations are rare and have little interference in the prediction. With our highly accurate kernel performance models, we can precisely predict the aggregation of compute kernel time and subsequently the E2E time of these models.

\section{Application and Discussion}
\label{sec:application_discussion}
\subsection{Case Study: Fast Sharding Config Selection Using Performance Modeling}
Industrial DLRM models can take days to train. Therefore, selecting a sharding configuration (i.e., the way to distribute embedding tables to multiple GPUs) that balances the loads on GPUs and speeds up the E2E per-iteration time is critical to reducing their training costs. Industrial sharding configs might consist of a sharding algorithm (greedy, multi-cost greedy, etc.), cost functions (multi-cost, memory-based, compute-based, etc.), table partition (column-based or row-based), and memory placement of tables (HBM/UVM). \JDO{It would be great to cite a sharding survey paper here.} \zhongyi{Honestly I don't have it~\ldots{} (Jason asked for it too.) I wrote this based on what I have seen inside Meta. Maybe we can leave this question to Louis.} Previously, the best configs were selected by benchmarking and grid-searching over a big search space formed by these factors, which can take as long as one day per workload. We consider using our performance model for this task so that \emph{without running the workload}, the selection time can be shortened to around 1 minute, with E2E per-iteration time predicted in seconds for each config. Notice again that we are \emph{not}, like some previous works did~\cite{Zha:2022:AAE,Zha:2022:DGE,Sethi:2022:RSF}, proposing a smart sharding algorithm here to achieve optimal performance. Instead, the goal is to quickly evaluate various sharding algorithms or configs and pick the fastest one for a specific problem size \emph{without benchmarking the model}.

\begin{table}
  \caption{Sharders and their indexing (i) or cost (c) functions, where $x$ represents an embedding table. An indexing function assigns a table directly to a certain rank, while a cost function estimates the cost of a table for the greedy algorithm.}
  \label{table:sharders}
  \centering
      \begin{tabular}{lc}
        \toprule
        Sharders & Functions \\
        \midrule
        naive & (i) $x.\textit{idx}\,\%\,\textit{ngpus}$ \\
        random & (i) $\textit{random}(\textit{ngpus})$ \\
        size\_greedy & (c) $x.E$ \\
        lookup\_greedy & (c) $x.L \cdot x.D$ \\
        norm\_lookup\_greedy & (c) $x.L / x.E$ \\
        size\_lookup\_greedy & (c) $x.L \cdot x.D \cdot \log_{10}(x.E)$ \\
        \bottomrule
      \end{tabular}
\end{table}

We conduct an experiment to demonstrate how our performance model can quickly select the best sharding config for DLRM training on a multi-GPU system. To show our idea with simplicity, we only consider sharding algorithms as the config and omit all other factors mentioned above. We also exclude recently sophisticated sharders~\cite{Zha:2022:AAE,Zha:2022:DGE,Sethi:2022:RSF}, although it is straightforward to integrate and test them in the future from an engineering point of view. We use six sharders listed in Table~\ref{table:sharders} for the experiment. In addition, we pick the 10 heavy tasks from the 20 tasks on each device generated in Section~\ref{sec:multi_gpu_e2e_tests} and run them again with the batch size 4096 and all sharders except for size\_lookup\_greedy. This is to guarantee that embedding lookup latency dominates the E2E time and thus the sharding config is likely to make a difference.

\begin{table*}
    \caption{Embedding table sharding config selection experiment results. Abbreviations: \textbf{P}: predicted; \textbf{A}: actual; \textbf{AP}: actual (time) of predicted (fastest config). Abs error is given by $\lvert \textit{Time}_{AP} - \textit{Time}_{A} \rvert / \textit{Time}_{A} \times 100\%$, which is 0 when the selected fastest config is exactly the actual fastest config. Notice again that Task\_x on different platforms are different workloads.}
    \label{table:config_selection}
    \centering
    \setlength{\tabcolsep}{3pt}
    \begin{small}
        \begin{tabular*}{\textwidth}{cc|ccccccc}
        \toprule
        Platforms & Tasks & Fastest (P) & Time (P, us) & Time (AP, us) & Fastest (A) & Time (A, us) & Abs Error & Meets Criterion? \\
        \midrule
        \multirow{10}{*}{4xGV100} & Task 0 & naive & 13.40 & 15.80 & size\_lookup\_greedy & 15.39 & 2.66\% & \cmark \\
        & Task 1 & naive & 12.13 & 15.46 & size\_lookup\_greedy & 13.59 & 13.76\% & \xmark \\
        & Task 2 & naive & 12.40 & 15.81 & random & 14.74 & 7.26\% & \cmark \\
        & Task 3 & size\_lookup\_greedy & 13.13 & 15.25 & random & 14.47 & 5.39\% & \cmark \\
        & Task 4 & norm\_lookup\_greedy & 12.26 & 15.54 & lookup\_greedy & 14.00 & 11.00\% & \xmark \\
        & Task 10 & naive & 11.77 & 15.13 & naive & 15.13 & 0.00\% & \cmark \\
        & Task 11 & naive & 11.75 & 14.22 & naive & 14.22 & 0.00\% & \cmark \\
        & Task 12 & naive & 12.39 & 15.07 & naive & 15.07 & 0.00\% & \cmark \\
        & Task 13 & norm\_lookup\_greedy & 12.61 & 16.04 & size\_lookup\_greedy & 14.96 & 7.22\% & \cmark \\
        & Task 14 & naive & 11.87 & 14.46 & lookup\_greedy & 13.13 & 10.13\% & \xmark \\
        \midrule
        \multirow{10}{*}{4xA100} & Task 0 & naive & 8.68 & 10.62 & naive & 10.62 & 0.00\% & \cmark \\
        & Task 1 & lookup\_greedy & 13.15 & 15.05 & random & 14.52 & 3.65\% & \cmark \\
        & Task 2 & naive & 9.81 & 11.43 & naive & 11.43 & 0.00\% & \cmark \\
        & Task 3 & naive & 11.31 & 14.04 & random & 13.85 & 1.37\% & \cmark \\
        & Task 4 & naive & 8.64 & 10.10 & naive & 10.10 & 0.00\% & \cmark \\
        & Task 10 & random & 8.69 & 10.85 & size\_lookup\_greedy & 10.38 & 4.53\% & \cmark \\
        & Task 11 & naive & 9.18 & 11.40 & naive & 11.40 & 0.00\% & \cmark \\
        & Task 12 & naive & 9.26 & 11.76 & size\_greedy & 11.54 & 1.91\% & \cmark \\
        & Task 13 & naive & 8.08 & 10.32 & size\_lookup\_greedy & 9.79 & 5.41\% & \cmark \\
        & Task 14 & naive & 8.13 & 9.67 & norm\_lookup\_greedy & 9.18 & 5.34\% & \cmark \\
        \bottomrule
        \end{tabular*}
    \end{small}
    \vskip -0.1in
\end{table*}

Table~\ref{table:config_selection} shows the actual and prediction time of using different sharding configs in selected tasks trained with platforms 4xGV100 and 4xA100. We set the success criterion to be \textbf{either the performance model accurately selects the fastest config, \emph{or} the absolute error between the \emph{actual} time of the predicted fastest config and the time of the actual fastest config is less than 10\%}. The reason for using this criterion is that we not only care about whether the fastest config is selected but also how close \emph{the actual time of the predicted fastest config} is from the actual fastest time when the fastest config is \emph{not} selected. This is because in practice failing to select the fastest config is tolerable as long as the actual time of the selected config is close enough to the actual fastest time. In the rightmost column, we see that the prediction result of our performance model meets the criterion in 17 out of 20 (85\%) tasks. In the 3 remaining tasks, the absolute errors are only 13.76\% at most. This demonstrates that our performance model is accurate enough to aid multi-GPU training optimization with low time and compute cost. Since we have demonstrated that our performance model has low prediction errors both generally and in individual cases, we are confident that it can also perform well in unseen future cases.


\subsection{Discussion}
\label{sec:discussion}
The biggest advantage of our performance model is its strong adaptability to new ML models. It is essentially an \emph{execution simulator} built in the granularity of kernels and ops. Supporting new ML models is straightforward, requiring only adding the missing performance models and overhead statistics of \emph{any} new kernels/ops. The library of supported kernels/ops grows as the system evolves, making it even easier to support new models later. More importantly, from the performance perspective, how the workload is \emph{executed} is more fundamental than how it is constructed because the same ML model can be trained with different strategies---e.g., data-parallel (DP), DDP, model-parallel (MP), pipelining, etc.---with completely different performance. However, our performance model can still handle all these cases because the easy-to-obtain execution trace (the input to our performance model) has information on both the model architecture and the execution. This means our performance model will extend well to unseen future workloads and execution paradigms.

Currently, this work has the following limitations:
\begin{itemize}
    \item Our prediction pipeline only supports ML workloads in FP32 precision. However, it can be seamlessly adapted for workloads in other precision types (FP16, INT8, etc.) by preparing kernel performance models for all kernels/ops with these types.
    \item This work currently only covers single-node multi-GPU performance modeling. To extend it to multi-node multi-GPU platforms, kernel performance models of \op{all-to-all} and \op{all-reduce} for the multi-node communication network must be prepared. Algorithm~\ref{alg:perf_model} should also be slightly modified to track multiple processes from all nodes. This will further adapt the performance prediction pipeline to the industrial environment for enormous ML workloads such as large language models (LLM)\@.
    \item The training data loading from data centers, usually impacted by uncontrollable factors like network speed and data center setups, might cause load imbalance in the industrial environment and is not considered in this work.
    \item Some additional infrastructure features, such as the support of capturing dynamic tensor sizes and fused ops information in the execution trace, can increase the robustness of this work on various types of ML workloads (such as training NLP models with variable-input-length and no padding) and cooperate with modern ML compilation and optimization techniques.
\end{itemize}
We plan to extend our code base to support these features.

\section{Conclusion and Future Work}
\label{sec:conclusion}
We extend our previous work on single-GPU performance model with two critical components for more accurate single- and multiple-GPU performance modeling: performance modeling of communication collectives and E2E training time prediction with inter-rank and intra-rank synchronization to enable multi-GPU training performance modeling of ML workloads, and support for input-data-distribution-aware performance modeling of embedding lookup as a general enhancement. We achieve high prediction accuracy on various types of ML workloads such as randomly generated DLRM and NLP models, and also demonstrate the performance model's ability to speed up DLRM training through the use case of embedding table sharding config selection.

There are a series of future works. Beyond the issues identified in Section~\ref{sec:discussion}, the data-distribution-aware method we use to model the performance of FBGEMM embedding lookup can be generalized to other sparse ops such as SpMM and SpGEMM to handle the training and inference performance prediction of future sparsified neural networks.

\section*{Acknowledgement}
\label{sec:acknowledgement}
We sincerely thank Daochen Zha, Valentin Andrei, and Yufei Zhu for their constructive feedback on this work.

\bibliographystyle{IEEEtran}
\bibliography{performance_model_paper}

\section*{Biography}
\label{sec:biography}
\begin{IEEEbiography}[{\includegraphics[width=1in,height=1.25in,clip,keepaspectratio]{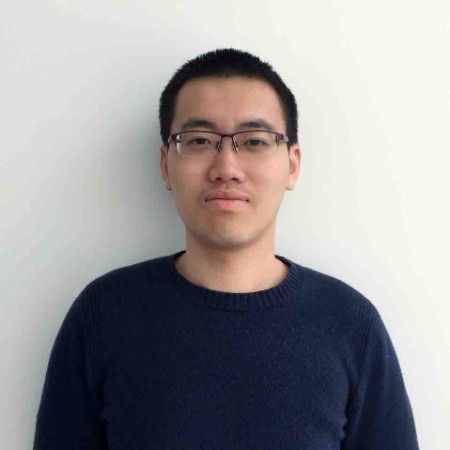}}] {Zhongyi Lin} is Senior Software Engineer at AMD, responsible for inference optimization of Large Language Models on specialized hardware for artificial intelligence. He is experienced in building high-performance machine-learning software systems for efficient training, inference, and performance modeling of language, recommendation, and vision models on devices such as GPU and AI chips. Prior to joining AMD, he received his Ph.D. degree in Electrical and Computer Engineering from University of California, Davis.
\end{IEEEbiography}

\begin{IEEEbiography}[{\includegraphics[width=1in,height=1.25in,clip,keepaspectratio]{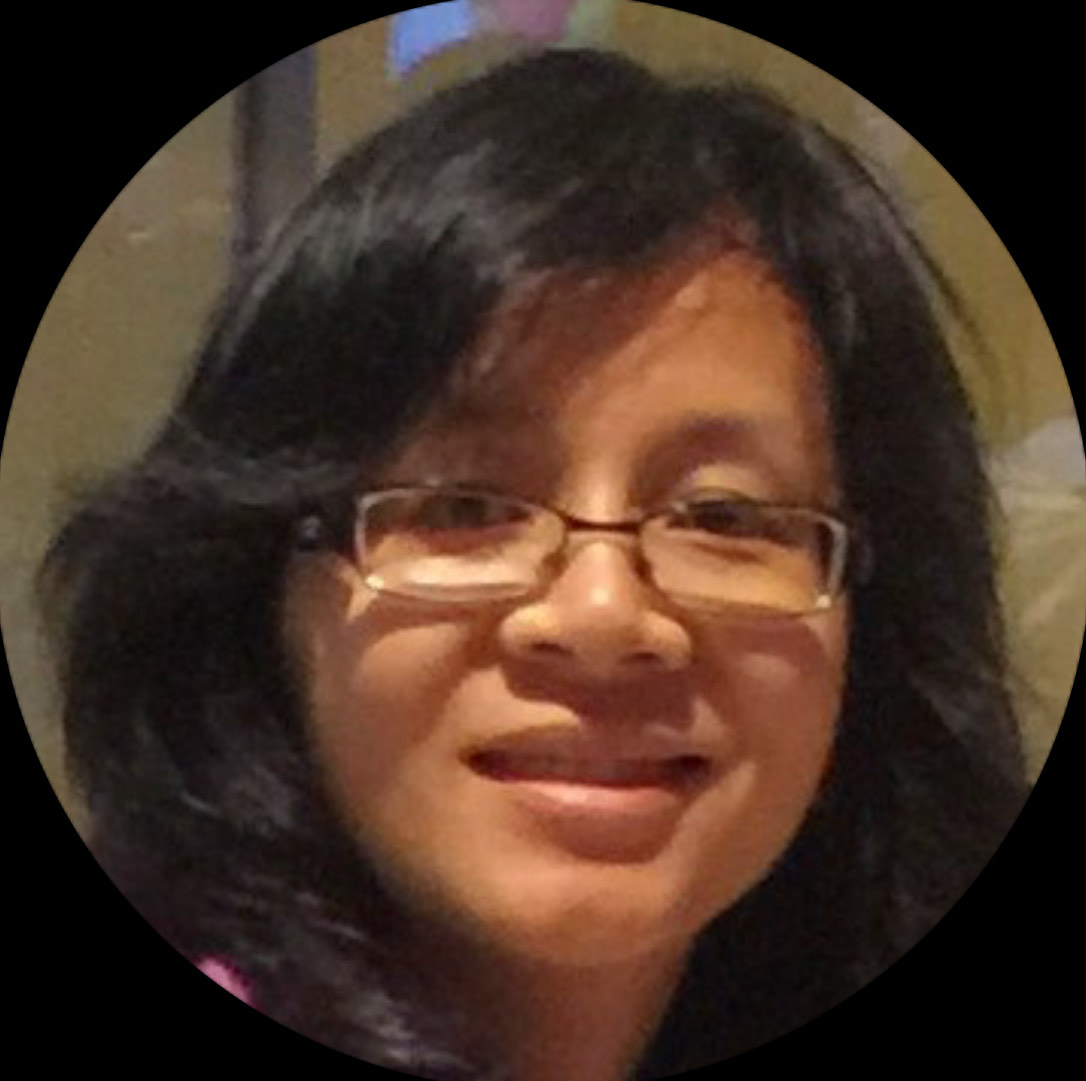}}] {Ning Sun} is a Senior Performance engineer at Meta. She has over 20 years of experience in developing tools to analyze performance and detect performance anomalies in very large distributed applications. At Meta, she is working on scaling AI training stack and HHVM server on multiple generations of hardware. Prior to joining Meta, she was a lead Contributor for SPECweb and SPECjAppServer while working in the Performance team at Sun/Oracle.
\end{IEEEbiography}

\begin{IEEEbiography}[{\includegraphics[width=1in,height=1.25in,clip,keepaspectratio]{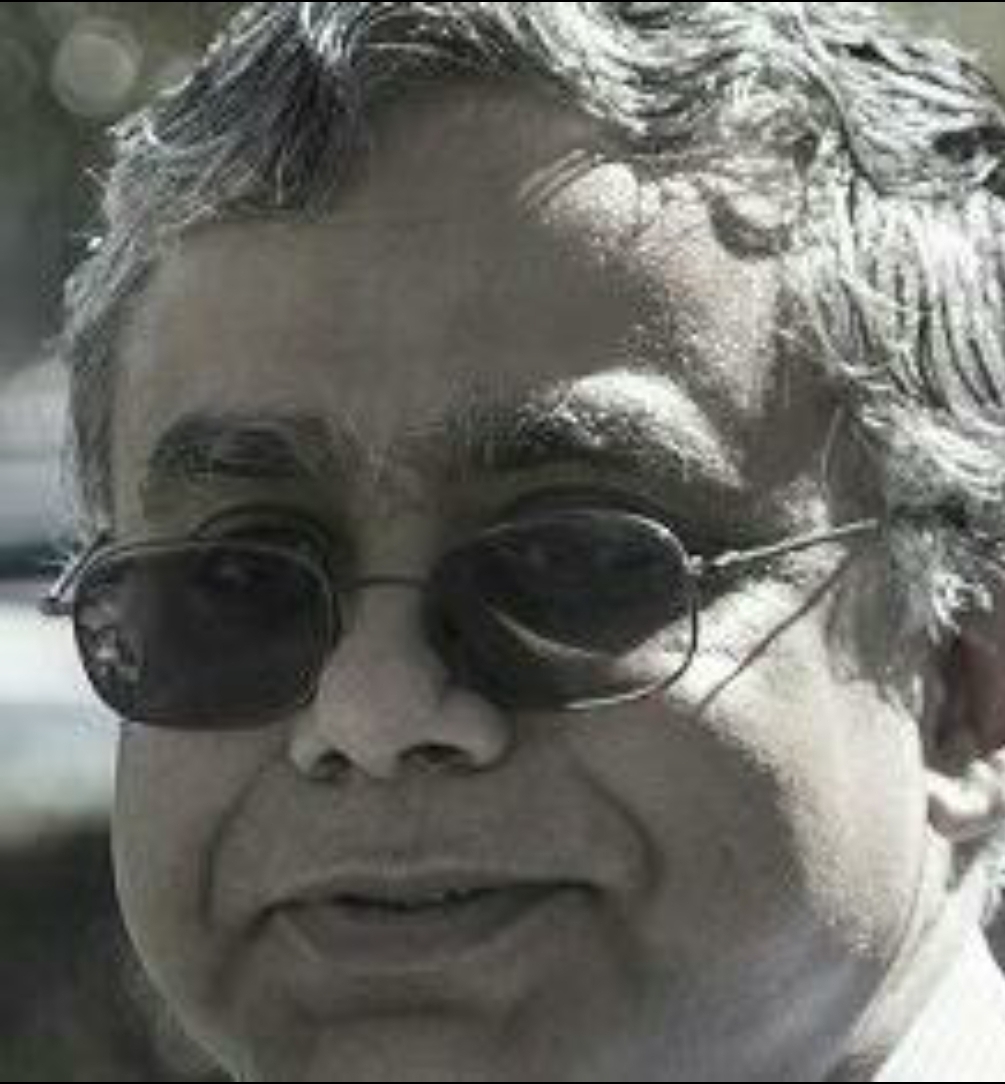}}] {Pallab Bhattacharya} is a Distinguished Engineer at Nvidia. Prior to joining Nvidia, he helped design and deploy AI Training Cluster at Meta, moving applications from CPU-based training to GPU-based training. He has extensive experience in High-Performance Communications using Infiniband and RDMA, building, debugging, and doing performance analysis of large-scale distributed and multi-thread applications. 
\end{IEEEbiography}

\begin{IEEEbiography}[{\includegraphics[width=1in,height=1.25in,clip,keepaspectratio]{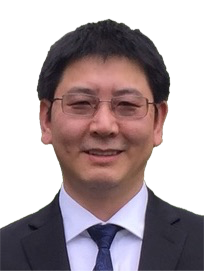}}] {Xizhou Feng} received his Ph.D. degree in Computer Science from the University of South Carolina and his J.D. degree from Marquette Law School. He is currently a software engineer at Meta Platform, Inc. His research interests include model stability and training efficiency of large AI models, scalable parallel algorithms, complex system modeling, high-performance computing, and the interactions between technology and law.
\end{IEEEbiography}

\begin{IEEEbiography}[{\includegraphics[width=1in,height=1.25in,clip,keepaspectratio]{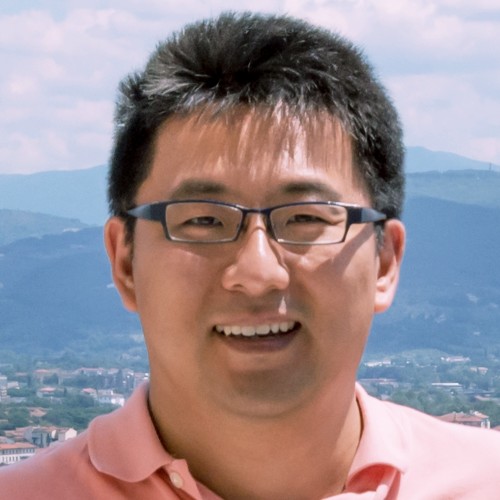}}] {Louis Feng} is an Engineering Manager at Meta, where he spearheads a team focused on improving fleet-wide efficiency visibility. His role involves a wide spectrum of responsibilities, including telemetry, performance analysis, and modeling. Louis’s notable contributions to the open-source community include his work on PyTorch execution trace and PARAM benchmarks, which have been adopted by the MLCommon’s Chakra working group. Before his tenure at Meta, Louis served as a staff performance engineer at Intel. There, he played key roles in the development of DreamWorks MoonRay, Pixar’s RenderMan, the Embree ray tracing engine, and the nGraph AI compiler framework. Louis received 12 patents and published papers at conferences like SIGGRAPH, NeurIPS, KDD, and ISCA. His contributions have sparked innovation across various industries.
\end{IEEEbiography}

\begin{IEEEbiography}[{\includegraphics[width=1in,height=1.25in,clip,keepaspectratio]{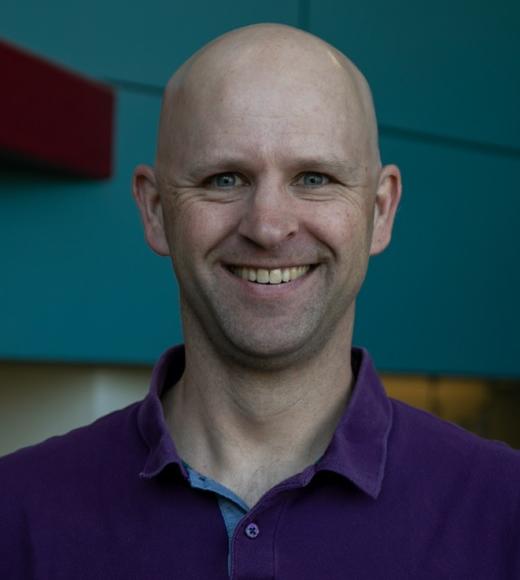}}]{John Owens} is the Child Family Professor of Engineering and  Entrepreneurship in the Department of Electrical and Computer Engineering at the University of California, Davis, where he leads a research group with a focus on GPU computing. He is an IEEE and AAAS Fellow and a Distinguished Member of the  ACM. John earned his Ph.D. in electrical engineering in 2003 from Stanford University and his B.S. in electrical engineering and computer sciences in 1995 from the University of California, Berkeley.
\end{IEEEbiography}

\end{document}